\documentclass[fleqn,usenatbib]{mnras}

\usepackage{newtxtext,newtxmath}
\usepackage{upgreek}
\usepackage{soul}

\usepackage{fix-cm}

\usepackage[T1]{fontenc}

\DeclareRobustCommand{\VAN}[3]{#2}
\let\VANthebibliography\thebibliography
\def\thebibliography{\DeclareRobustCommand{\VAN}[3]{##3}\VANthebibliography}

\usepackage{graphicx}	
\usepackage{amsmath}	

\usepackage{caption}

\title[Bar properties as a function of wavelength in TNG50]{Bar properties as a function of wavelength in TNG50: analysis of mock images}

\author[G. F. Gonçalves et al.]{
Gustavo F. Gonçalves$^{1}$\thanks{E-mail: goncalvesg@alunos.utfpr.edu.br},
Rubens E. G. Machado$^{1}$, Karín Menéndez-Delmestre$^{2}$, Thiago Bueno-Dalpiaz$^{2}$
\\
$^{1}$Departamento Acadêmico de Física, Universidade Tecnológica Federal do Paraná, Av. Sete de Setembro 3165, Curitiba 80230-901, PR, Brazil\\
$^{2}$Universidade Federal do Rio de Janeiro, Observatório do Valongo, Ladeira Pedro Antônio, 43, Saúde CEP 20080-090 Rio de Janeiro, RJ, Brazil
}

\date{Accepted 2025 August 14. Received 2025 July 27; in original form 2025 March 25}

\pubyear{\the\year{}}

\begin{document}
\label{firstpage}
\pagerange{\pageref{firstpage}--\pageref{lastpage}}
\maketitle

\begin{abstract}
Recent studies used TNG50 galaxies to study bar formation, evolution, and properties like length, strength, and pattern speed. In simulations, these are typically derived from particle positions and mass distribution, neglecting stellar light and extinction effects. However, observational studies indicate that bar appearance depends on wavelength. To test whether this dependence exists in TNG50 at \( z \approx 0 \), we analysed 50 strongly-barred galaxies using mock images from SKIRT radiative transfer simulations covering infrared to ultraviolet filters (\textit{Spitzer} $3.6 \, \upmu \text{m}$, SDSS \textit{i}, \textit{r}, \textit{g}, S-PLUS \textit{J0378}, \textit{GALEX NUV}, and \textit{GALEX FUV}). Bar ellipticity and length were measured via ellipse fitting. Ellipticity generally increases by 6 per cent from $3.6 \, \upmu \text{m}$ to $g$ band, and by 9 per cent to $J0378$ band. On average, TNG50 bars cannot be said to be longer in bluer filters when the entire sample is used. However, the trend is indeed detected when only star-forming galaxies are considered. In this star-forming subsample, bar length increases by 10 per cent from $3.6~\mu \text{m}$ to $g$ band, and by 17 per cent to $J0378$ band; moreover, the bar can appear up to 20--30 per cent longer in bluer mocks than in the mass map. Over 90 per cent of bars vanish in UV due to minimal emission by dominant stellar populations. We reproduced the bar properties morphological dependence phenomenon using age-filtered mass maps, where older stars form shorter, rounder bars, and younger stars generate longer, more elliptical ones. TNG50 bars exhibit a wavelength-dependent trend similar to observations: bars more elliptical and longer in bluer filters, with this effect being stronger in star-forming galaxies.

\end{abstract}




\section{Introduction}
    \label{sec:intro}

Galactic bars are important and common structures in the central regions of spiral galaxies. Their presence, size, and strength play a crucial role in shaping the evolution of the galactic disc by redistributing angular momentum among the various components of the galaxy \citep{2003Athanassoula_angular}. The fraction of bars in the local universe has been extensively studied \citep[e.g.][]{1963Vaucoulerus_bar_frac, 2007Karin_bar_frac, 2010Nair_bar_frac, 2018Erwin_bardfrac_local}, with strong and weak bars observed in approximately two-thirds of all spiral galaxies. Beyond the local universe, recent studies with the JWST indicate that bars are observable at higher redshift ($1 \leq z \leq 3$), although they are found in a smaller fraction of galaxies \citep{2023Costantin_redshift3_mw_like, 2024LeConte_bars_z3, 2024Guo_JWSTbarfrac}.

Properties of observational systems are a key factor influencing the identification or non-identification of bars in galaxies. Spatial resolution limitations can also hinder bar detection \citep[e.g.]{2007Karin_bar_frac}, as well as the wavelength range studied, with optical surveys typically identifying a smaller fraction of bars compared to infrared observations \citep{2000Eskridge_barfrac_infrared}. Additionally, bar strength impacts detection; weak and mildly elliptical bars are generally more challenging to identify \citep{2003Sheth_weakbars_identific}.

A recent observational paper by \citet{2024Karin_wavebandependence} investigates the behaviour of measuring galactic bars in the local universe across different observational filters, ranging from infrared to ultraviolet (\textit{Galex NUV}, \textit{Galex FUV}, Johnson \textit{B}, Johnson \textit{R}, \textit{Spitzer} IRAC/$3.6 \, \upmu \text{m}$). The aim of \citet{2024Karin_wavebandependence} is to understand the effect of wavelength-dependent characteristics of the same bar to extrapolate findings to band-shifting scenarios at high redshift. The methodology is based on measuring bar properties by fitting ellipses \citep{1987Jedrzejewski_karin_ellipse_method}, where the maximum ellipticity determines bar length. The main result that the authors present is the wavelength dependence of bar length and strength (characterized by ellipticity) with observed band: they find that in bluer filters bands tend to show bars that are more elongated and elliptical (by around 10 per cent), while the same bar appears shorter and less elliptical in infrared. Hypotheses to explain this dependency have been proposed, including the possible influence of star-forming clumps at the bar ends or the effect of kinematic fractionation \citep{2017Athanassoula_kinematic_MW_bulge, 2017Fragkoud_kinematic_frac_bulge, 2017Debattista_bar_kinematic_pop, 2020Neumann_bar_pop}, which results in younger stars being preferentially located along the length of the bar, leading to an elongation in bluer filters. Additionally, the contribution of a bulge in redder filters could reduce bar ellipticity, affecting its isophotes \citep{2024Karin_wavebandependence, 2016Dias_ellipse_use}. 

Isolating such phenomena and testing these hypotheses remains challenging in observations. In this context, we propose using cosmological simulations to explore this phenomenon. Our primary objective is to investigate whether this fine wavelength-dependent detail is captured in cosmological simulations, as it remains unclear if simulations reproduce this phenomenon; to date, no publications have addressed this aspect.

Hydrodynamic \textit{N}-body simulations are widely used to model the formation and evolution of galactic bars. One approach to studying bar properties is through simulations of isolated galaxies, where bars are known to form spontaneously due to disc instability over a few Gyr. The timescale of bar formation and the strength of the resulting bar are influenced by various properties of the disc and halo, including the central concentration of the halo \citep{2002Athanassoula_halofraction_bar, 2003Athanassoula_barslowdown_halo}, the shape of the halo \citep{2010Machado_haloshape, 2015Iannuzzi_haloshape}, and the gas fraction in the disc \citep{2013Athanassoula_barevo_length}. Interactions between bars and classical bulges in isolated galaxy scenarios are discussed in \citet{2012Saha_spinbulges, 2016Saha_spinbulge, 2024Machado_classical_bulges}. Examples of interactions between bars and galactic warps are presented in \citet{2018Zana_flyby_bar_warp, 2024Andressa_warp}. However, bars are also strongly influenced by their environment, where tidal interactions can induce bar formation \citep{2018Lokas_tidalinduced, 2019Peschken_tidally_Illustris}, and environmental interactions can lead to bar destruction \citep{2024Rosas-Guevara_barfall_environment}. In this context, large-scale cosmological simulations become essential tools.

Cosmological simulations play an important role in studying the formation and evolution of the fraction of barred galaxies. The properties and evolution of bars have been explored using data from simulations such as EAGLE \citep{2017Algorry_Eagle_barform,2022Eagle_barevo} and IllustrisTNG, in both TNG100 \citep{2020Zhao_barredTNG100} and TNG50 \citep{2022RosasGuevara_BarredCat}. Notably, when comparing the bar fractions from simulations with observational results, there is a good agreement at redshift \(z = 0\). However, this fraction tends to diverge at higher redshift: while simulations like TNG50 show a bar fraction close to 40 per cent \citep{2022RosasGuevara_BarredCat} at redshift \(z = 3\), recent observational results from JWST indicate a significant decrease in this fraction \citep{2024Guo_JWSTbarfrac}.

In works using IllustrisTNG, the properties of galaxies are often evaluated through the positions, kinematics and mass distribution of stellar particles \citep[e.g.][]{2022Zana_morphologicalTNG, 2022RosasGuevara_BarredCat, 2020Zhao_barredTNG100, 2024Semczuk_patternspeed, 2022Izquierdo_bardiscinstability}, since the mass distribution is closely correlated with light emission, this is valid when disregarding effects of stellar populations and extinction effects on morphology, bringing everything closer to a mass-based model. However, for detailed analyses involving wavelength-dependent phenomena, it is necessary to go beyond mass-based models. Accurate characterization of morphological features requires simulations that take into account radiative transfer and dust extinction effects derived from the simulation gas, allowing for a more accurate representation of the emission and scattering of light by stars and interstellar matter.

Recent studies have utilized galaxies from cosmological simulations to create mock images that simulate observational filters. Two notable codes used for this purpose are GALAXEV \citep{2011Bruzual_Galaxev} and SKIRT~\citep{2011Baes_oldSKIRTrelease}. GALAXEV is a stellar population synthesis code with low computational cost; however, it has limitations in representing star-forming regions and does not account for radiative transfer (RT) or extinction effects. In contrast, the SKIRT code employs Monte Carlo techniques for radiative simulations, enabling detailed modelling of star-forming regions and incorporating attenuation due to dust extinction in gas-derived models. For our purposes, we will use RT simulations using SKIRT due to its more complete physics for regions of star formation and the presence of dust in the central region of galaxies.

To provide context, we can refer to works that employ mock images based on cosmological simulations. \citet{2024Baes_AtlasTNGSKIRT_method} created an atlas of TNG50 mock galaxies at various wavelengths, each with different orientations relative to the cosmological cube. This atlas, generated using SKIRT, features noise-free simulations. An example analysis included evaluating the effective radius of galaxies across different wavelengths \citep{2024Baes_AtlasTNGSKIRT_radius}. Another example of creating mock images of individual galaxies is presented in \citet{2019Rodriguez-Gomez_MocksTNGPanstars}, which uses TNG simulations processed with SKIRT \citep{2020Baes_SKIRT9release} to model observations from Pan-STARRS (Survey Telescope and Rapid Response System; \citealt{2016Chambers_Panstars}). These simulations include steps for adding synthetic noise, enhancing their resemblance to real observational data. A distinct application is presented in \citet{2023Snyder_lightcone_TNG}, which uses galaxies from TNG to simulate light cone mosaics analogous to those observed with JWST. By concatenating multiple galaxies into a single field of view, this approach models deep field observations across a series of broadband filters utilized by JWST-NIRCam and the Hubble Space Telescope cameras.

In this work, we constructed our own set of radiative transfer-based mock images using the SKIRT code, with source galaxies drawn from the TNG50 simulation. Our goal is to replicate the approach of \citet{2024Karin_wavebandependence}, as previously described, to assess whether the waveband dependence observed in local universe galaxies is also present in this simulation. The TNG50 simulation was chosen for this analysis due to its ability to replicate several galaxy properties in the local universe. These include its successful reproduction of mass-size relations and disc galaxy morphology, comparable to real galaxies \citep{2019Rodriguez-Gomez_MocksTNGPanstars}, and its good performance in multi-wavelength analyses such as the Tully-Fisher relation \citep{2025Baes_Tully_FisherTNG50}.

Among large cosmological simulations, TNG50 and TNG100 shows a bar fraction at \(z = 0\) that is most consistent with observations. This improvement over its predecessor, Illustris, is largely attributed to updates in the subgrid physics model in the IllustrisTNG suite, particularly in the treatment of AGN and stellar feedback \citep{2020Zhou_TNGvsIllustris_bars}. In contrast, the original Illustris \citep{2019Peschken_tidally_Illustris} and Eagle \citep{2017Algorry_Eagle_barform} simulations report bar fractions of around 20 per cent at \(z = 0\), and New Horizon \citep{2022Reddish_New_Horizon_nobars} shows no barred galaxies at this redshift. The Auriga simulation \citep{2025Fragkoudi_bar_frac} exhibits comparably good results for the bar fraction at \(z = 0\) and a better representation of bar fraction evolution. Due to its zoom-in nature, it focuses on 40 Milky Way-like halos, while TNG50 evaluates the bar fraction in hundreds of halos that formed spontaneously within a broader cosmological environment.

To ensure that our entire sample comprises barred galaxies, we began with the selection from \citet{2022RosasGuevara_BarredCat}. This catalogue guarantees that all chosen subhalos from TNG50 contain bars, identified by Fourier decomposition of the face-on stellar surface density. Specifically, we used the group of approximately 50 galaxies classified as strongly barred $\left(A_{2,\max} \geq 0.4\right)$. We excluded weakly barred galaxies to avoid potential detectability issues after convolution with an observational PSF (which will be the subject of a future study).

While TNG50 already has an atlas produced with SKIRT by \citet{2024Baes_AtlasTNGSKIRT_method} that includes our sample subhalos, we opted to generate our own files. This choice allowed us to align all galaxies in a face-on orientation according to the angular momentum of their discs, which was important for two main reasons: it enables direct comparison with measurements made by other authors who also used face-on orientations \citep[e.g][]{2022RosasGuevara_BarredCat, 2024Lu_ellipse_TNG50}, and it ensures that the effects of bar inclination do not interfere with our measurement results. Additionally, generating our own files allowed us to select specific observational filters and wavelengths and apply synthetic noise.

Our work is organized as follows. Section~\ref{sec:methods} provides details on the TNG50 simulation, from which the halos were selected as inputs for the RT simulation using SKIRT, including steps for noise addition and ellipse fitting measurements. Section~\ref{sec:results} presents the statistical results of our analysis on the sample of 50 galaxies. Section~\ref{sec:discussion} discusses the results obtained from our analysis, and Section~\ref{sec:conclusion} summarizes and presents our conclusions.

\section{Simulations and methods}
    \label{sec:methods}

This section presents the sample of barred galaxies in the TNG50 cosmological simulation, as well as the methods used to process images of the mocks created with SKIRT and measure the morphological properties of the bars.

\subsection{The IllustrisTNG Simulation}

The IllustrisTNG project \citep{2018Pillepich_TNGrelease, 2019Nelson_TNGpublicrelease} provides a suite of simulations that represent the current state-of-the-art of galaxy formation and evolution \citep{2018Pillepich_Galaxyformationmodel_TNG, 2017Weinberg_GalaxyformationmodelTNG} based on hydrodynamical cosmological models. It is performed using the moving-mesh code AREPO \citep{2010Springel_AREPOrelease} considering a flat $\Lambda$CDM cosmological model, containing additional physics algorithms in addition to the gravo-magneto-hydrodynamical code such as star formation, colling, feedback from supernovae and active galactic nuclei. The IllustrisTNG simulation executed runs within three cubic cosmological volumes, each having side lengths of 300, 100, and 50 cMpc, respectively, with each volume featuring progressively finer spatial resolutions. This setup allows not only the formation of large cosmological structures, but also the study of subhalos containing galaxies in detail, enabling morphological analyses of their components. Examples of such analyses can be found in the works of \citet{2019Tacchella_morphologyTNG} and \citet{2022Zana_morphologicalTNG}.

Individual galaxies in IllustrisTNG are catalogued by assigning an ID in each snapshot, which is updated as the redshift evolves in subsequent snapshots. Both dark matter and baryonic components are grouped into subhalos using a Friend-of-Friend algorithm \citep{2001Springel_sublink}, and these subhalos are linked to specific families of IDs with progenitor and descendant ID across different redshifts using the SUBLINK algorithm \citep{2015Rodriguez_SUBLINK}.

The TNG50 simulation from IllustrisTNG provides the highest resolution among the TNG suite, making it suitable to explore sub-galactic components. The simulation features a dark matter (DM) mass resolution of \(4.5 \times 10^5\) \({\rm M_{\odot}}\) and a mean baryonic mass resolution of \(8.5 \times 10^4\) \({\rm M_{\odot}}\). The box size is 51.7 Mpc, and the number of particles is \(2 \times 2160^3\). The minimum gas cell softening length is 72 parsecs, while the gravitational softening length for dark matter and stars starts at 0.288 kpc at \(z=0\), gradually increasing to 0.58 kpc at higher redshifts and reducing to 0.29 kpc at lower redshifts. TNG50 has been the focus of various studies on barred galaxies and their properties \citep[e.g.][]{2024Semczuk_patternspeed, 2022Frankel_TNG_barpattern, 2022Izquierdo_bardiscinstability, 2022RosasGuevara_BarredCat}.
 
\citet{2022RosasGuevara_BarredCat} undertook a detailed analysis in TNG50 galaxies to identify bar structures in galaxies at $z \sim$ 0~--~4. The authors employ an automated approach to ensure robust detection of bars in galaxies at different stages. The results of this work culminated in the creation of a supplementary data catalogue, which has been made available to the scientific community via the Data Access section of the IllustrisTNG project.\footnote{\hyperlink{http://www.tng-project.org/data/docs/specifications/}{http://www.tng-project.org/data/docs/specifications/}} This catalog not only enhances our understanding of bar dynamics over time but also serves as a  essential  resource for comparative studies.

We specifically selected the list of strong bars from \citet{2022RosasGuevara_BarredCat}, focusing on barred galaxies at \(z = 0\). This targeted selection was essential for the initial sample used in our analyses. The galaxies in the sample have stellar masses greater than \(10^{10}\rm M_{\odot}\). At \(z = 0\), 30 per cent of the disc galaxies are barred, with approximately 11 per cent classified as strongly barred based on Fourier decomposition methods. The total number of disc galaxies at \(z = 0\) is 349, of which 105 are barred, and 53 are strongly barred $\left(A_{2,\max} \geq 0.4\right)$. Our final sample is comprised by the strongly-barred galaxies from \citet{2022RosasGuevara_BarredCat}.

\subsection{Generating mock images}

Previous studies have primarily evaluated the properties of galaxies based on the positions, kinematics, and mass distribution of stellar particles in the IllustrisTNG simulation \citep[e.g.][]{2020Zhao_barredTNG100, 2022Zana_morphologicalTNG, 2022RosasGuevara_BarredCat, 2022Izquierdo_bardiscinstability, 2024Semczuk_patternspeed}. However, to effectively compare observed morphological characteristics of stellar structures, it is essential to analyse the behaviour of the light emitted by galaxies. Since the emission and scattering of light by stars and interstellar matter are crucial components of this analysis \citep{2010Gadotti_lightdust_morphology, 2010Baes_lightdust_morphology}, the use of mock images that simulate light emission is necessary. These simulations must incorporate stellar properties and account for the effects of interstellar dust, enabling a more precise and detailed analysis.

The radiative transfer (RT) simulation stage, which considers the stars of the TNG50 galaxies as the source and simulates the dust properties generated from the gas in the subhalos, was performed using the Monte Carlo RT code SKIRT \citep{2011Baes_oldSKIRTrelease, 2020Baes_SKIRT9release}. SKIRT calculates physical processes, including scattering, absorption, and emission through the transfer medium \citep{2015Baes_SKIRTphysics&functions}, and features an import interface compatible with data formats from  hydrodynamical simulations like IllustrisTNG. This allows the import of stellar information as radiative sources and the derivation of the spatial distribution and properties of dust from the gas particles. In recent years, SKIRT has been consistently used for RT calculations with dust absorption, using particles from the TNG50 simulation as input \citep[e.g.][]{2019Rodriguez-Gomez_MocksTNGPanstars, 2024Baes_AtlasTNGSKIRT_radius, 2024Baes_AtlasTNGSKIRT_method, 2022Popping_SKIRTandTNG}.

The steps for creating our pipeline to run simulations with SKIRT are based on \citet{2022Trcka_pipelineSKIRT}. Many of the parameter choices were made following their procedures, and we refer to this source for further details.

The primary radiation sources of each galaxy are generated from the information of the stars, which are divided into two stellar groups: those older than 10 Myr and the star-forming regions. The older stars are modelled using the \citet{2003Bruzual_SEDoldstars} SED library, which is suitable for representing mature stellar populations due to its detailed treatment of stellar evolution. These stellar populations utilize a \citet{2003Chabrier_massfuction} initial mass function. The star-forming regions are represented by the MAPPINGS-III library \citep{2008Grooves_SEDstarforming}, which captures the complex interactions between young stars and the interstellar medium, including the effects of ionization and dust. These choices ensure a more accurate simulation of the different stellar populations within the galaxies.

The medium composed of a diffuse dust disc is derived from the properties of the gas particles in the TNG50 galaxies. The ISM model used is THEMIS \citep{2017Thesmis_dustmodel}, and the steps to separate the interstellar medium gas from the hot circumgalactic gas follow the threshold (equation \ref{eq:ISM_Torrey}) proposed by \citet{2022Trcka_pipelineSKIRT}, based on the gas temperature and density relationship described by \citet{2019Torrey_ISMgasTNG}:

\begin{equation}
    \log\left(\frac{T_{\text{gas}}}{\text{K}}\right) = 6 + 0{.}25 \log\left(\frac{\rho_{\text{gas}}}{10^{10} \, h^2 \, \text{kpc}^{-3}}\right)
	\label{eq:ISM_Torrey}
\end{equation}

\begin{figure}
	\includegraphics[]{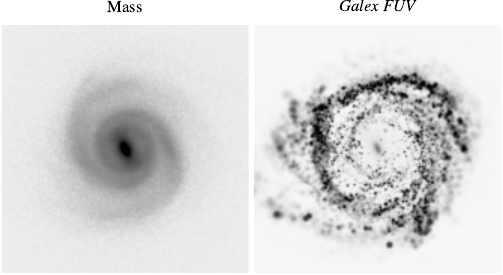}
    \caption{Comparison between the stellar mass map and the mock \textit{GALEX NUV} image in their idealized noise-free stage for the barred galaxy subhalo ID394621. Each panel shows a face-on view with a field of 60 × 60 kpc.}
    \label{fig:Fig1}
\end{figure}

\begin{figure*}
	\includegraphics[]{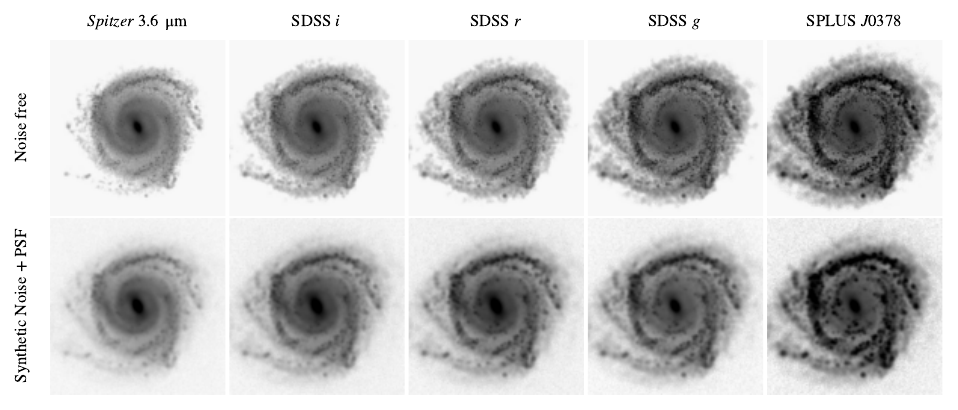}
    \caption{Comparison of the stellar mass maps and mock images in the three observational bands considered in the waveband dependence analysis. The top row shows the five bands in their idealized noise-free stage. The bottom row shows the respective versions of the bands after the addition of synthetic noise and PSF. The barred galaxy used in this example is the subhalo ID394621. All the figures were generated with a size of 60 $\times$ 60 kpc in each frame and with a face-on orientation. For additional ultraviolet views of galaxy ID394621 and further examples of galaxies analysed in this study, see Appendix \ref{fig:appendix1_A}.}
    \label{fig:mosaic_FITS}
\end{figure*}

Using stellar particle data and dust distribution derived from the gas component of the TNG50 galaxies, data cubes were generated with bands covering the entire wavelength range from 0.1 to 5 $\mu$m from the RT simulations. The images generated by SKIRT are saved in FITS format (Flexible Image Transport System). By using the Python toolkit for working with SKIRT (PTS) library \citep{2020Baes_SKIRT9release}, it is possible to generate images in specific observational bands by applying appropriate transmission curves. For our analysis, we generated images in the in the infrared using the \textit{Spitzer} IRAC/$3.6 \, \upmu \text{m}$ channel, along with SDSS \textit{i}, \textit{r}, \textit{g} filters, and the SPLUS narrowband \textit{J0378} (near SDSS \textit{u}). We also explored ultraviolet filters, using \textit{Galex FUV} and \textit{Galex NUV} (see Appendix \ref{fig:appendix1_A}). The standard SDSS \textit{u} band proved unsuitable for our analysis due to the clumpy appearance of its mocks from star-forming regions, which severely affected ellipse fitting. Similar issues were encountered with the GALEX FUV and NUV
filters discussed in Subsection \ref{sec:UV_bars}. Therefore, the SPLUS \textit{J0378} narrowband filter was used as a viable alternative to the problematic SDSS \textit{u} band, providing a shorter wavelength coverage than SDSS \textit{g} but without effects of excessive star-forming clumps.

The Southern Photometric Local Universe Survey (SPLUS) \citep{2019Oliveira_SPLUS} utilizes a system of 12 optical filters of the Javalambre system \citep{2019Cenarro_JPLUS_bands}: four broad bands --- \textit{u} (Javalambre-like), \textit{g}, \textit{r}, and \textit{i} (SDSS-like) --- and eight narrow bands, which are strategically distributed across the wavelength ranges of these broadband. This filter set is complementary to surveys like SDSS and allows for detailed studies of galaxy photometry and morphologies. Notably, the feasibility of simulating observations with SPLUS filters, by incorporating their transmission curves into radiative transfer codes like SKIRT, has already been successfully demonstrated in \citet{2024Castelli_SPLUS_SKIRT}. This established modelling capability supports our use of the SPLUS \textit{J0378} filter in our simulations.

Additionally, we produce projected stellar mass maps in FITS format, based on TNG50 star particle spatial information. These maps provide a direct representation of the stellar mass distribution in galaxies without additional radiative transfer processing. Throughout the paper, we refer to these stellar mass maps and the five observational bands as \textit{M}, $3.6 \, \upmu \text{m}$, \textit{i}, \textit{r}, \textit{g}, and \textit{J0378}, respectively. Fig.~\ref{fig:Fig1} and the top row of Fig.~\ref{fig:mosaic_FITS} show an example of these maps for one of the galaxies in our barred sample.

\subsection{Adding synthetic realism to mock images}

The simulated images generated by SKIRT are generally idealized without the presence of noise or any additional elements that would add observational realism to the image. As indicated in previous studies that created simulated images of galaxies from Illustris \citep{2014Vogelsberger_Illustirs1releasae} and IllustrisTNG \citep{2019Nelson_TNGpublicrelease} using RT \citep{2019Rodriguez-Gomez_MocksTNGPanstars, 2023Snyder_lightcone_TNG, 2017Bottrell_Illutris1mocks}, we implemented two crucial steps to add synthetic realism to the analysis: the inclusion of read noise and sky background noise as well as the convolution with a Point Spread Function (PSF).

We incorporate two types of noise that are commonly encountered in real astronomical observations. First, Gaussian noise is applied to simulate read noise, with its standard deviation set to the mean signal of the galaxy pixels divided by a target signal-to-noise ratio. Subsequently, Poisson noise is introduced to represent sky background noise, which is directly influenced by the flux present in each pixel. This combination of noise types is crucial for accurately modelling the typical noise found in astronomical data.

We performed a convolution using a PSF equivalent to a FWHM of $\sim$ 1 kpc which corresponds to $\sim$ 1 arcsec for galaxies at redshift $z~\sim~0.05$ (see Appendix \ref{fig:appendix_B} for further discussion on this PSF and the robustness of our results to its variations). This smoothing was applied through a Gaussian convolution algorithm available in the Astropy library \citep{2018_astropy1}. 

PSF convolution proved particularly useful in resolving a systematic issue in our measurements of TNG50 bars. As noted in \citet{2024Goncalves_PSF}, the isophotes in the central regions of simulated bars in TNG50 tend to be excessively elongated, unlike the morphology typically observed in real galaxies. This excessive central elongation alters the expected ellipticity profile, potentially affecting the interpretation of bar properties. However, by incorporating realistic noise elements, such as PSF convolution, sky background, and read noise, we mitigate these distortions, bringing the isophotal contours into better agreement with observations.

\subsection{Ellipse fitting and bar properties}

\begin{figure}
	\includegraphics[]{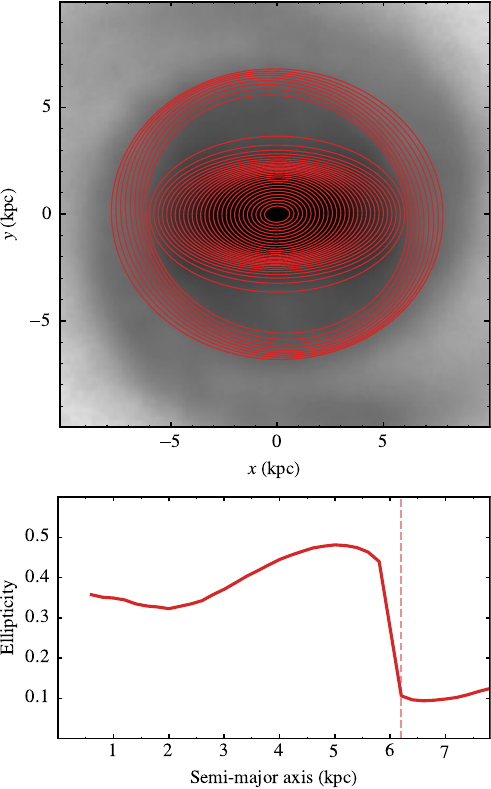}
    \caption{Method for estimating the length and ellipticity of the bar. Top panel shows the spatial properties of the bar estimated by fitting ellipses to isophotes in the central region of the galaxy using mass map. Bottom panel presents the ellipticity of the ellipses as a function of the semi-major axis, with the dashed line marking the length of the bar estimated using the ellipticity drop method after the identified maximum. The barred galaxy used in this example is the subhalo ID394621 with a face-on orientation and rotated so that the semi-major axis of the bar is parallel to the x-axis.}
    \label{fig:ellipse_fitting}
\end{figure}

Various methods are employed to measure the length and ellipticity of the bar, both in observations and in simulated galaxies. Among the most commonly used methods are Fourier analysis, as exemplified in the study by \citet{2022RosasGuevara_BarredCat} on TNG50 galaxies, and the fitting of ellipses to isodensity curves\citep{1987Jedrzejewski_karin_ellipse_method,1995Wozniak_ellipsefitmethod, 2005Erwin_ellipsefitmethod}, as utilized in \citet{2020Zhao_barredTNG100} and \citet{2024Lu_ellipse_TNG50}, also with IllustrisTNG data. Fourier analysis is effective in measuring bars by decomposing the light or projected mass distribution into symmetric components, while ellipse fitting is widely used to characterize the shape of the bar based on the spatial distribution of isodensities, providing a detailed measure of ellipticity and length. A comparative analysis of these methods has been explored in observational and numerical contexts, as discussed in \citet{2006Dansac_ellipse_A2_comparison}, highlighting their strengths and limitations in different regimes.

Ellipse fitting is one of the most widely used approaches for studying bar properties, as it provides a fast and computationally efficient method suitable for large samples. Compared to more complex techniques such as IMFIT \citep{2015Erwin_infit}, GALFIT \citep{2002Peng_galfit}, and BUDDA \citep{2004Souza_budda}, which require higher-quality data and longer processing times, ellipse fitting remains a practical and robust choice in various studies \citep[e.g.][]{2007Karin_bar_frac, 2008Sheth_ellipse_fitting_use, 2016Dias_ellipse_use}. In this work, we implement this method using the astropy Photutils Isophote package \citep{2016Bradley_photutilsAstropy}, an updated and improved version of the ELLIPSE task from IRAF\footnote{IRAF is distributed by the National Optical Astronomy Observatory, which is operated by the Association of Universities for Research in Astronomy (AURA) under a cooperative agreement with the National Science Foundation.} \citep{1987Jedrzejewski_karin_ellipse_method}. We used a constant and linear step size of 0.25 pixels and during the fitting, both the ellipticity and position angle were allowed to vary, while the centre of the ellipse was fixed at the galaxy centre. Our choice also ensures methodological consistency with the recent study by \citet{2024Karin_wavebandependence}, facilitating a direct comparison between results.

The length of the bar can be defined based on the properties of the ellipses fitted to the isophotes of the central region of the galaxy. However, the criteria used by different authors to delimit the outer boundary of the bar and characterize its transition into the rest of the galactic disc can vary. Typically, a clearly defined bar is expected to exhibit a monotonic increase in ellipticity along its length, reaching a peak before experiencing a sharp drop, which marks the transition to the galactic disc \citep[e.g.][]{2007Karin_bar_frac}. This behaviour is frequently used as a signature of bars in observational studies. However, deviations from this pattern are commonly noted. 

Barred galaxies of TNG50 tend to exhibit excessively high central ellipticities, with inner isodensity contours presenting very elongated shapes \citep{2024Goncalves_PSF, 2024Lu_ellipse_TNG50}. This contrasts with observations that often show rounder isophotes in the galaxy centre region.  This discrepancy might stem from the absence of nuclear discs within TNG50 bars, potentially due to resolution limitations for structures smaller than 1 kpc, constrained by the numerical resolution of discs and softening length. Observational studies, analysing sub-kpc resolution structures, such as The TIMER project by \citet{2019Gadotti_TIMER} which surveyed nearby barred galaxies with VLT MUSE \citep{2010Bacon_MUSE}, indicate that most bars in observed galaxies possess nuclear discs \citep{2020Gadotti_nuclear_disks}, which remains a known challenge to reproduce in simulations.

To circumvent certain measurement pitfalls, such as the high ellipticity in the central region and the presence of bars with ellipticity plateaus, we chose to adopt the methodology of \citet{2020Zhao_barredTNG100}. This method defines the end of the bar as the region where the ellipticity drops by 15 per cent from its maximum value, meaning the bar is considered to extend over the region that retains 85 per cent of the maximum ellipticity (denoted $R_{85}$). The bar measurement methodology can be seen in Fig.~\ref{fig:ellipse_fitting}, where the upper panel illustrates the fitted ellipses and the sharp change in ellipticity at the end of the bar, and the lower panel shows the ellipticity versus semi-major axis plot. 

During the exercise of fitting concentric ellipses and measuring the bar in galaxies, a few difficulties were identified that interrupted the unsupervised flow of measurements. The main issue, which occurred before the addition of synthetic noise to the mocks, was the increase in ellipticity toward the centres of galaxies. This feature, noted in simulation results \citep[e.g.][]{1990Athanassoula_highellipticalcenter, 2020Zhao_barredTNG100, 2021Fragkoudi_ellipticitycenter}, was resolved by convolving with a PSF. In \citet{2024Goncalves_PSF}, we presented an example of the ellipticity profile of a simulated barred galaxy from TNG50, measured with and without smoothing. The inclusion of a PSF convolution was shown to render the inner ellipticity more comparable to the typical behaviour of observed galaxies. Additionally, we noted that bluer observational bands, where clumps and knots of star formation are frequent, prevented the convergence of ellipse fittings in certain situations. These star formation clumps and knots, also reported in the observational analysis of \citet{2024Karin_wavebandependence}, located not in the bars but in adjacent regions of the galactic disc, were addressed with segmentation masks. Such masks, also applied in the present paper, rectified the flux of these areas with typical values from the surroundings in the disc, enabling the convergence of ellipse fitting. Out of the original sample of 53 galaxies, the ellipse fitting converged properly for 50 of them. This is the sample to be used in the remainder of the paper.

\section{Results}
    \label{sec:results}

In this section, we analyse the wavelength dependence of bar properties in our simulated galaxies. We examine how bar ellipticity and length vary across different mock observation filters, identifying key trends and significant dependencies within specific galaxy subsamples. We also explore the implications for bar detectability in the UV mocks and correlation between these morphological dependencies and stellar age distributions. A detailed explanation of the methodology used to compare the bar length measurements across filters is provided in Sections \ref{sec:3.1} and \ref{sec:3.2}. The physical mechanisms associated with the host galaxy that help explain these systematic differences are discussed in Section \ref{sec:3.3}.

\subsection{Comparison of wavebands across individual galaxies}
    \label{sec:3.1}

\begin{figure}
	\includegraphics[]{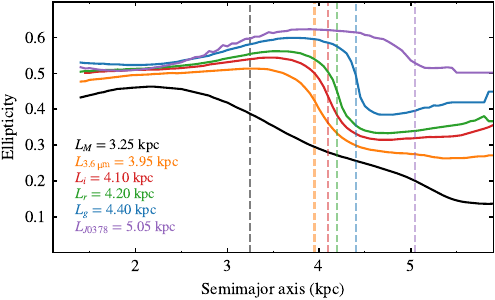}
    \caption{Results of the ellipse-fit on the isophotes of the observational bands analyzed. The solid lines represent Mass (black), \textit{Spitzer} $3.6 \, \upmu \text{m}$ (orange), SDSS \textit{r} (red), SDSS \textit{i} (green), SDSS \textit{g} (blue) and SPLUS \textit{J0378} (purple). The colored dashed lines identify the respective estimate of the bar length in each band. The barred galaxy used in this example is the subhalo ID566365.}
    \label{fig:ellipse_fitting_filters}
\end{figure}

We applied the bar measurement method to all 50 galaxies in our sample, considering the \textit{J0378}, \textit{g}, \textit{r}, \textit{i} bands, the \textit{Spitzer} IRAC/$3.6 \, \upmu \text{m}$, and the mass maps. For each galaxy, we obtained a comparison profile of the ellipticity across the different filters, which defines both the bar length and its maximum ellipticity. An example of this result is shown in Fig.~\ref{fig:ellipse_fitting_filters}, where we present an example of the ellipse fit on the isophotes of a barred galaxy (subhalo ID566365), analysing the ellipticity profiles in the 2D mass maps and mocks generated using the five observational filters. The solid lines in Fig.~\ref{fig:ellipse_fitting_filters} show the ellipticity variations of the fitted isophotes as a function of the semi-major axis for each filter, with the dashed lines indicating the estimated bar length in each band. The bar lengths vary between 3.25\,kpc for the mass, 3.95\,kpc for \textit{Spitzer} $3.6 \, \upmu \text{m}$, 4.10\,kpc for \textit{i-}, 4.20\,kpc for \textit{r-}, 4.40 kpc for \textit{g-} and 5.05 kpc for \textit{J038-}band. 

\begin{figure}
	\includegraphics[]{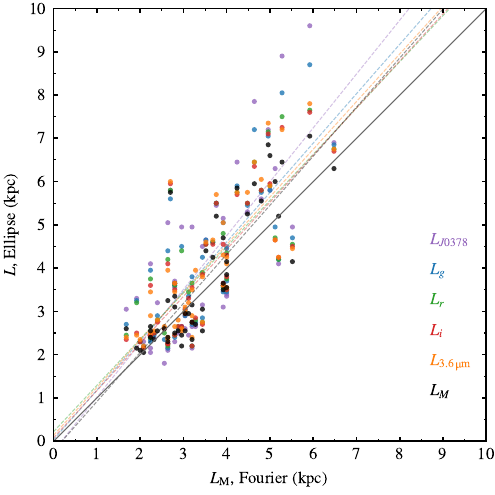}
    \caption{Comparison between bar length measurements using the Fourier decomposition method from \citet{2022RosasGuevara_BarredCat} and the ellipse fitting method (this work). The purple, blue, green, red, orange and black symbols correspond to the bar length measured in the \textit{J0378-, g-, r-, i-}bands, the IRAC/$3.6 \, \upmu \text{m}$ and mass maps respectively. The solid line represents the one-to-one relation. Dashed lines of matching colour indicate linear fits to the data points in each band.}
    \label{fig:fourier_ellipse_comp}
\end{figure}

Comparison between bar length measurements using the Fourier decomposition method from \citet{2022RosasGuevara_BarredCat} and the ellipse fitting method (this work) can be found in Fig.~\ref{fig:fourier_ellipse_comp}. The distribution of points reveals that, for short bars, the Fourier method tends to produce values both above and below those obtained via ellipse fitting. However, for longer bars ($\ge4$ kpc), the Fourier method systematically yields smaller measurements, this is evident in both the mass maps and when compared to observational filters. Linear fits (represented by dashed lines in the same colour as the data points in Fig.~\ref{fig:fourier_ellipse_comp}) emphasizes this trend, on average, all filters show longer bar lengths when measured with the ellipse fitting method compared to Fourier analysis. This offset is most prominent in the bluer filters (\textit{J0378} and \textit{g-}band), while redder filters and the mass maps yield bar length measurements that are more consistent among themselves.

Current evidence \citep{2024Karin_wavebandependence} suggests that the differences in bar measurements across observation bands arise not from a characteristic of the measurement itself, but from a physical and morphological phenomenon within the bar. This phenomenon varies among galaxies, causing the dependence on the observation band to be more pronounced in some cases, while absent in others. Thus, a statistical analysis is necessary to describe the potential dependencies within the sample. Two main hypotheses have been proposed to explain this effect: (1) the influence of bulge size and star-forming knots at the ends of the bar, and (2) the presence of young stellar populations along the bar due to kinematic fractionation.

\begin{figure*}
	\includegraphics[]{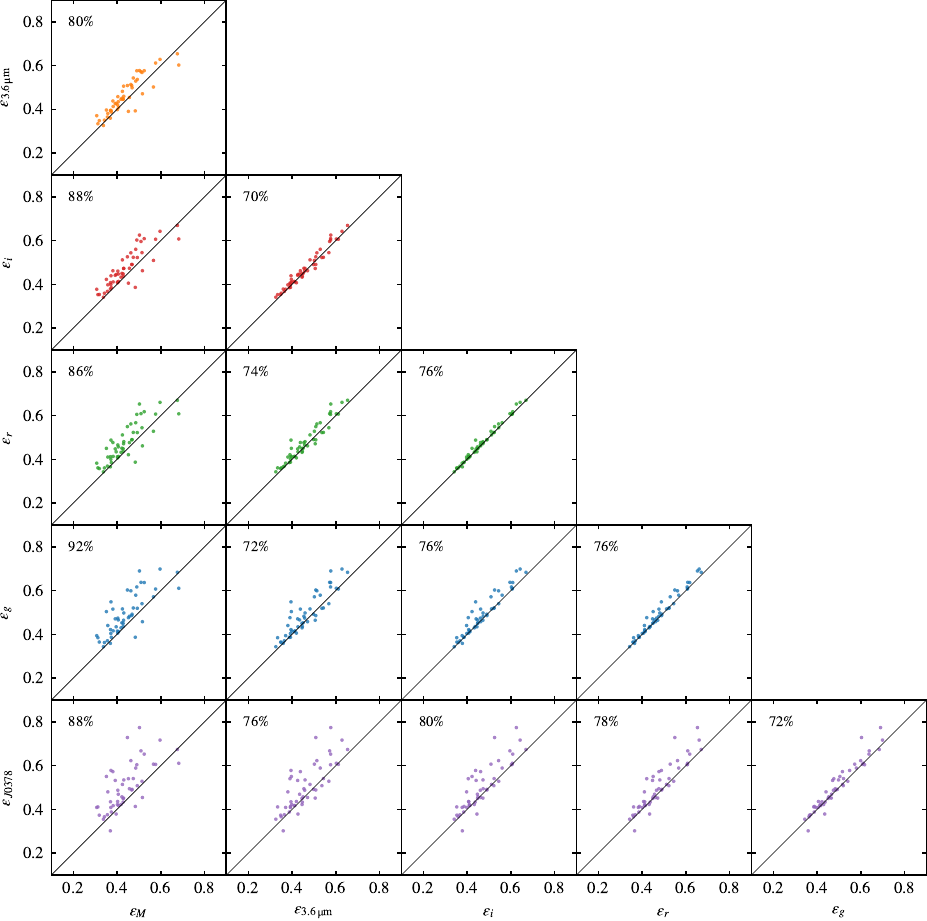}
    \caption{Measurements of the bar ellipticity of all the galaxies in our sample. Each frame corresponds to a comparison between two observational bands, where the bluer band is always on the y-axis of the comparison. The percentage in the top left corner of each frame indicates the fraction of points above the identity line.}
    \label{fig:corner_plot_e_L}
\end{figure*}

After measuring the ellipticity profiles for each galaxy in 5 filters and mass map, we obtained arrays containing 50 measurements of bar length (defined by $R_{85}$)  and ellipticity (defined by the maximum of the ellipticity profile) for each waveband. These arrays enabled us to perform a filter-by-filter comparison. Initially, we compared the individual values for each galaxy, as shown in the identity line corner plots in Fig.~\ref{fig:corner_plot_e_L}. We also compared the average values across filters and applied the Wilcoxon test to assess the statistical significance of the differences between the samples.

The comparison between filters in the corner plots of Fig.~\ref{fig:corner_plot_e_L} follows an arbitrarily established arrangement, where the measurement from the bluer filter (with shorter wavelength) is plotted on the y-axis, while the redder filter is on the x-axis. This setup is intended to facilitate the visualization of potential variations between the filters. Significant differences in the measurements are identified by calculating the ratio of galaxies above the identity line, given by the equation:

\begin{equation}
    f_{\text{identity}} = \frac{N_{\text{galaxies above identity line}}}{N_{\text{total galaxies}}}.
	\label{eq:ratio1}
\end{equation}

\begin{figure*}
	\includegraphics[]{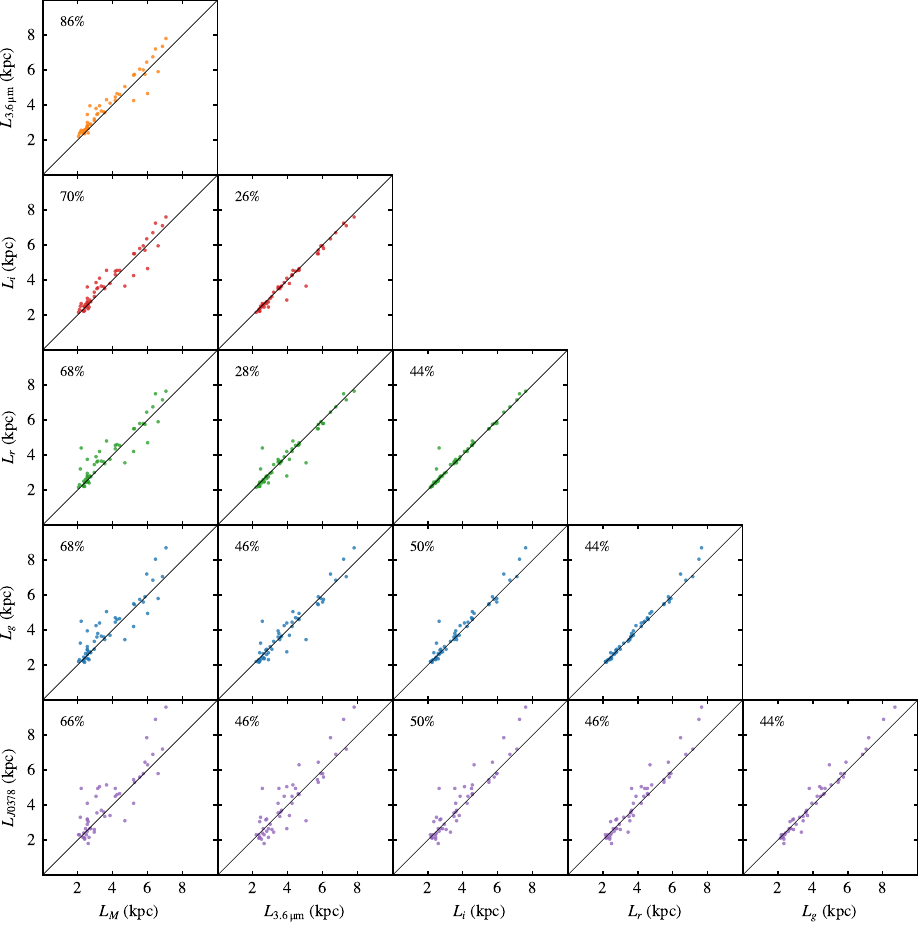}
    \caption{Same as Fig.~\ref{fig:corner_plot_e_L} but for bar length measurements instead of ellipticity. Each frame corresponds to a comparison between two observational bands, where the bluer band is always on the y-axis of the comparison. The percentage in the top left corner of each frame indicates the fraction of points above the identity line.}
    \label{fig:corner_plot_L}
\end{figure*}

This ratio is  consistently high in all cases analysed for ellipticity (Fig.~\ref{fig:corner_plot_e_L}). However, the bar length measurements are initially inconclusive (Fig.~\ref{fig:corner_plot_L}). For most filter comparisons, the ratio of galaxies with lengths above the identity line to the total number of galaxies remains close to 50 per cent. Notably, comparisons between redder filters, which are less affected by ongoing star formation, consistently show a low ratio of galaxies above the identity line. This apparent lack of wavelength dependence in the overall sample will become clear once the galaxies are analysed as a function of star formation rate.

\begin{figure*}
	\includegraphics[]{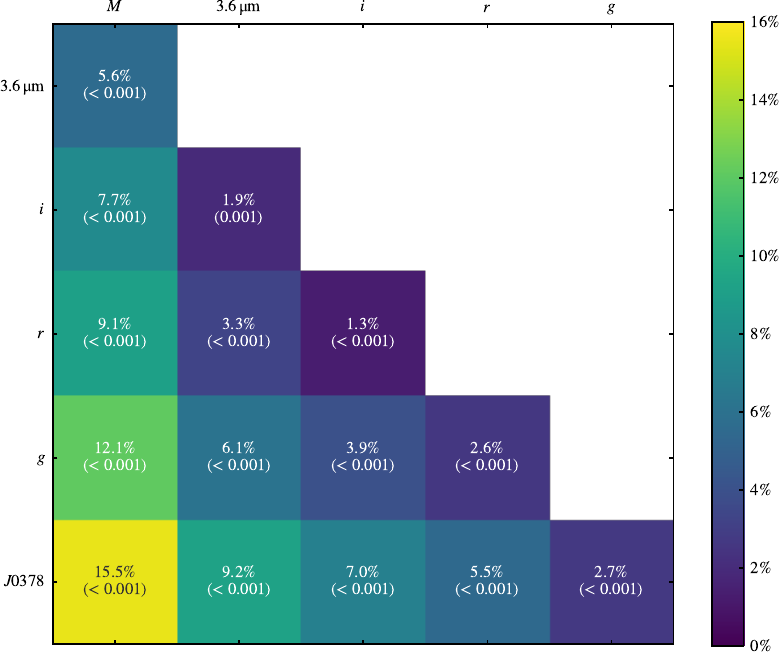}
    \caption{Comparison of the bar ellipticity across different bands. The percentage indicates how much greater the ellipticity was on average when comparing the filters. The significance of each measurement was tested using the Wilcoxon test, with the results shown in parentheses.}
    \label{fig:meanratio_E}
\end{figure*}

\begin{figure*}
	\includegraphics[]{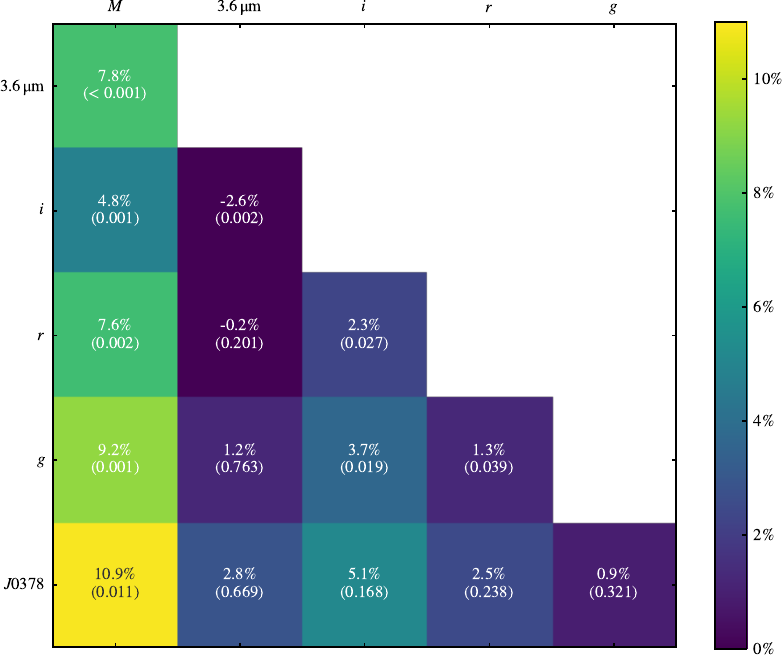}
    \caption{Comparison of the bar length across different bands. The percentage indicates how much greater the length was on average when comparing the filters. The significance of each measurement was tested using the Wilcoxon test, with the results shown in parentheses.}
    \label{fig:meanratio_L}
\end{figure*}

Another approach to analysing the differences in bar length and ellipticity measurements is to calculate the ratios between pairs of measurements for all six combinations of our wavebands. The ratio \( f_{\text{pair}} \) is defined as:

\begin{equation}
    f_{\text{pair}} = \frac{L_{\text{bluer}} - L_{\text{redder}}}{L_{\text{redder}}},
	\label{eq:ratio2}
\end{equation}

\noindent where the average value of \( f_{\text{pair}} \) represents the percentage by which the comparison between the pair of filters is calculated for the 50 galaxies. The significance of this measure is evaluated using the Wilcoxon test, with the results presented in Fig.~\ref{fig:meanratio_E} for ellipticity and Fig.~\ref{fig:meanratio_L} for length.

The plot comparing the $f_{\text{pair}}$ values for the ellipticities across different filters, shown in Fig.~\ref{fig:meanratio_E}, reveals a clear and consistent trend. The measured ellipticity differences are gradually larger in bluer filters, both when comparing to the stellar mass map and when comparing observational filters among themselves. This trend holds without exception, indicating that the filter covering the shortest wavelength consistently exhibits higher ellipticity.

In the plot comparing bar length measurements, the mean difference between filters is initially small and lacks statistical significance, as indicated by the values in parentheses in Fig.~\ref{fig:meanratio_L}. However, this does not imply that bar lengths are approximately equal across all filters. Instead, we will see that this apparently inconclusive result is an effect caused by two distinct SFR populations.

In this work, the Wilcoxon test is used to assess whether the differences in bar ellipticity and length measurements, obtained from different observational filters, are statistically significant. The choice of the Wilcoxon test is motivated by the fact that the measurements from different filters do not necessarily follow a normal distribution and may exhibit subtle variations between filters. Moreover, since we are dealing with paired samples—i.e., the same galaxies observed at multiple wavelengths—the Wilcoxon test is particularly suitable for comparing these differences robustly, without assuming a specific distribution of the data.

Specifically, by applying the Wilcoxon test, we compare the values of bar ellipticity and length between the filters for each galaxy in the sample, aiming to determine whether the noted differences can be explained by random statistical variations or if they reflect a real dependence between the bands. The use of this test allows us to rigorously verify whether, for example, bluer filters consistently yield measurements that differ from those of redder filters. Thus, the test is used not only to confirm the presence of variations in the bar properties but also to ensure that these differences are statistically significant, as indicated by the $p$-values provided in parentheses in Fig.~\ref{fig:meanratio_E}, Fig.~\ref{fig:meanratio_L} and Fig.~\ref{fig:meanratio_L_sSFR11}. These values help us determine the statistical significance of the differences, with values less than 0.05 being considered sufficient to reject the null hypothesis that the samples have identical means, implying that the differences are indeed statistically meaningful.

\subsection{Monotonicity test between wavelengths}
    \label{sec:3.2}

To assess whether the differences in measurements identified in the previous section exhibit any monotonic dependence between the analysed filters, we undertake a series of tests. The tests used to evaluate monotonicity and the slope of this trend are Spearman's Rank Correlation Test and a linear slope fitting test. In other words, we aim to quantify whether the variations in bar length and ellipticity across filters follow a systematic trend with wavelength.

The use of Spearman's test in this context involves comparing the bar ellipticity and length measurements by fixing the sequence of observational filters along the $x$-axis in order of increasing wavelength. Measurements derived from the 2D mass maps were not included in the Spearman's test. We begin with the \textit{Spitzer} IRAC/$3.6 \, \upmu \text{m}$ followed by SDSS \textit{i}, \textit{r}, \textit{g} and ending with the shortest wavelength filter, SPLUS \textit{J0378}. This arrangement allows us to evaluate the potential relationship between the measurements and the variation in wavelength. 

\begin{figure}
	\includegraphics[]{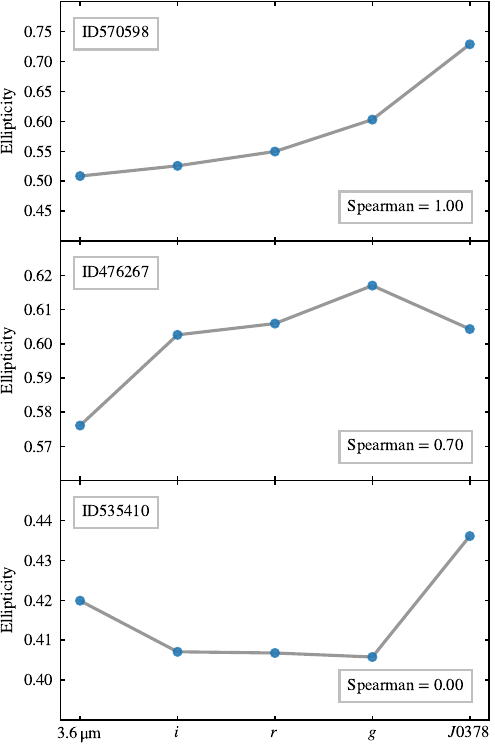}
    \caption{Illustration of the Spearman test applied to assess whether the relationship between increasing ellipticity across different observational bands is monotonic, quasi-monotonic, or non-monotonic, using the bars of selected example galaxies. The top panel show an example of a monotonic relationship. The middle panel illustrates a quasi-monotonic relationship. The bottom panel presents a case where non-monotonic trend is identified.}
    \label{fig:spearman_test}
\end{figure}

Our initial hypothesis is that the ellipticity measurements will show a monotonic increase as we progress to filters with shorter wavelengths (from the $3.6 \, \upmu \text{m}$ to the observational band \textit{J0378}). However, as illustrated in Fig. \ref{fig:spearman_test}, where we display the ellipticities measured in the $3.6 \, \upmu \text{m}$, \textit{i}-, \textit{r}-, \textit{g}- and \textit{J0378}-bands for 3 galaxies in our sample, the trend with band can vary. When the trend is strictly postiive monotonic, the Spearman coefficient will be equal to 1.00, indicating a perfect correlation. In cases where the monotonicity is weaker or absent, the coefficient will be smaller than 1.00, reflecting a quasi-monotonic or non-monotonic trend. For the purpose of Spearman's test, the filter names are treated as dimensionless integers along the $x$-axis. It is important to note that equivalent examples to these described for a strong positive trend also exist for a strong negative monotonic trend. A Spearman coefficient greater than 0.8 (or less than $-0.8$ for a negative trend) is typically considered to indicate a strong monotonic trend.

\begin{figure}
	\includegraphics[]{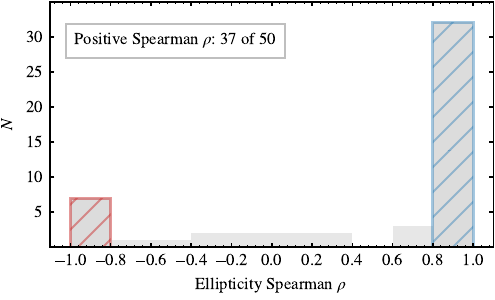}
    \caption{Histogram displaying the distribution of Spearman correlation coefficients, which test for monotonicity, for the ellipticity of bars across different bands. The blue‐hatched bin highlights galaxies with a strong positive trend, and the red‐hatched bin highlights galaxies with a strong negative trend --- both of which are discussed in detail in Fig.~\ref{fig:slope_e_sSFR}}
    \label{fig:hist_rho_ell}
\end{figure}

For ellipticity, the number of Spearman coefficients greater than 0.8 is 33 out of 50, indicating a perfect correlation in many cases. Considering the distribution of values for the entire sample, 75 per cent of the coefficients show a positive trend regarding wavelength dependence, suggesting a strong positive monotonicity between ellipticity measurements and the different observational bands. These results are displayed in Fig.~\ref{fig:hist_rho_ell}, which highlights the consistency of this dependence across the sample of galaxies. 

\begin{figure}
	\includegraphics[]{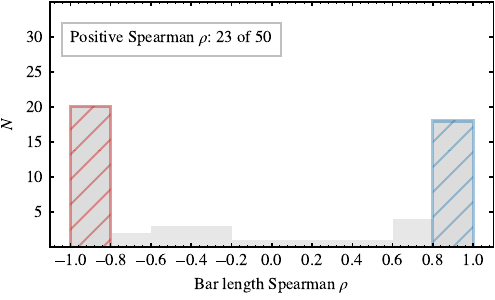}
    \caption{Histogram displaying the distribution of Spearman correlation coefficients, which test for monotonicity, for the lengths of bars across different bands.The blue‐hatched bin highlights galaxies with a strong positive trend, and the red‐hatched bin highlights galaxies with a strong negative trend --- both of which are discussed in detail in Fig.~\ref{fig:slope_L_L}}
    \label{fig:hist_rho_SMA}
\end{figure}

In contrast, a monotonic trend for bar length is detected in 38 out of 50 galaxies; however, this divides into 18 examples of high positive monotonicity and 20 examples of high negative monotonicity (Fig.~\ref{fig:hist_rho_SMA}). This bimodality will become
clear once the galaxies are analysed as a function of star formation
rate. Overall, a monotonic trend is present in both ellipticity and bar length measurements, but it is considerably stronger when evaluating ellipticity. These results suggest that the variation in the morphological properties of the bars is more sensitive to changes in observational bands for ellipticity than for length.

One limitation of Spearman's test is that, while it effectively measures the monotonicity of dependencies, it does not assess the rate at which one variable changes relative to another. Spearman's test, when applied to such a small number of data points, indicates whether a monotonic relationship exists—whether the measurements consistently increase or decrease—but it does not account for the steepness or magnitude of these changes, as a linear analysis would. Consequently, both small and large variations in measurements across different observational filters can yield the same Spearman coefficient, as the test only evaluates the order of the variations, not their magnitude.

To address this limitation and assess the magnitude of the variations in a continuous way, a linear fit is employed. The linear fit not only evaluates the direction of the trend but also quantifies the slope, representing the rate of change in the measurements as the observational filters vary. This method allows for the distinction between trends with small variations and those exhibiting large changes, providing a more comprehensive view of the relationship between the measurements and the filters. For the linear fit, the observational filters are fixed along the $x$-axis in the same increasing wavelength order used in Spearman's test: starting with the infrared filter and ending with the shortest wavelength filter. For the purpose of the linear fit, the filter names are also treated as dimensionless integers along the $x$-axis.

\begin{figure}
	\includegraphics[]{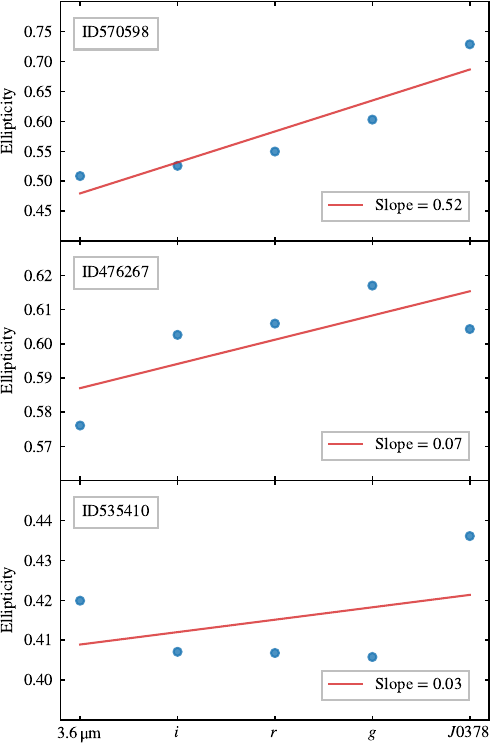}
    \caption{Illustration of the linear fit applied to quantify the dependence of ellipticity on different observational bands for the bars of the same three galaxies shown in Fig.~\ref{fig:spearman_test}. The top frame show an example quantifying a strong dependence. The middle frame illustrates a case quantifying a quasi-monotonic but weak dependence. The bottom frame presents an example quantifying a weak or negligible dependence. Note the different vertical scales.}
    \label{fig:slope_test}
\end{figure}

As an example of this approach, we utilize the same three galaxies show in Fig. \ref{fig:spearman_test}, to illustrate the Spearman's test. These examples illustrate different scenarios: one with a steep slope, indicating significant variation in the measurements across filters; one with an intermediate slope, where the changes are moderate; and one with a non-monotonic trend, where the variation is minimal. In this last case, while Spearman's test correctly identifies the absence of a monotonic relationship, the linear fit reveals that the variation between filters is small, even without a clear trend. This example is illustrated in Fig.~\ref{fig:slope_test}.

By combining Spearman's test, which measures monotonicity, with the linear fit, which evaluates the slope of the trend, a more robust analysis of the dependencies between ellipticity and bar length measurements across different observational filters is achieved. While Spearman's test provides insight into the order of the variations, the linear fit quantifies the strength of these variations, offering a more detailed understanding of the relationships between the measurements and the mass map and filters.

\subsection{Correlations with the host galaxy}
    \label{sec:3.3}

By simultaneously evaluating the results from Spearman's test and the slope of the linear fit, we can highlight that the positive dependence for the phenomenon under investigation is present in the majority of the galaxies in our sample. It is notably strong in 66 per cent of the cases for ellipticity, while being absent in 18 per cent of the 50 galaxies analysed for ellipticity. This indicates a general trend where the phenomenon is prevalent, though with varying degrees of strength across the sample.

Using the slope obtained from the linear fit, we compared it with several physical properties of the host galaxies, searching for correlations with properties such as galaxy mass, star formation rates, and the presence of stronger or more elliptical bars. Among these physical properties, the one that exhibited the strongest correlation with the phenomenon was the specific star formation rate (sSFR).

\begin{figure}
	\includegraphics[]{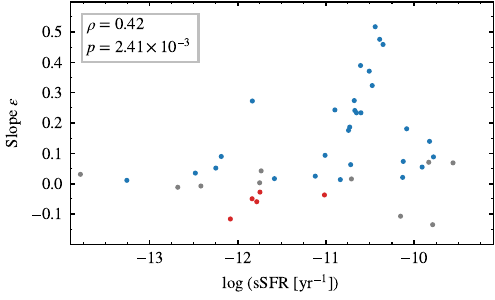}
    \caption{Correlation between specific star formation rate and the slope of bar ellipticity measured across bands. Each point represents one of the 50 galaxies in our sample. Galaxies with a strong positive monotonic trend in ellipticity are highlighted in blue. Galaxies with a strong negative monotonic trend in ellipticity are highlighted in red. The remaining galaxies are shown in gray. The Spearman correlation shown refers to the full sample.}
    \label{fig:slope_e_sSFR}
\end{figure}

As an illustration of this correlation, we plotted the slope of ellipticity as a function of the specific star formation rate of the host galaxies (Fig.~\ref{fig:slope_e_sSFR}). This analysis was performed for both the entire sample of 50 galaxies and a subset consisting only of galaxies with a Spearman coefficient greater than 0.8, indicating a strong monotonic dependence on wavelength. In both cases, the correlation suggests a clear trend: galaxies with higher star formation activity tend to exhibit a stronger wavelength-dependent effect, while galaxies with lower star formation rates are typically the ones showing non-monotonic behaviour and lower slopes. The phenomenon studied appears to manifest more notably in star-forming galaxies and is less present in passive galaxies.

\begin{figure}
	\includegraphics[]{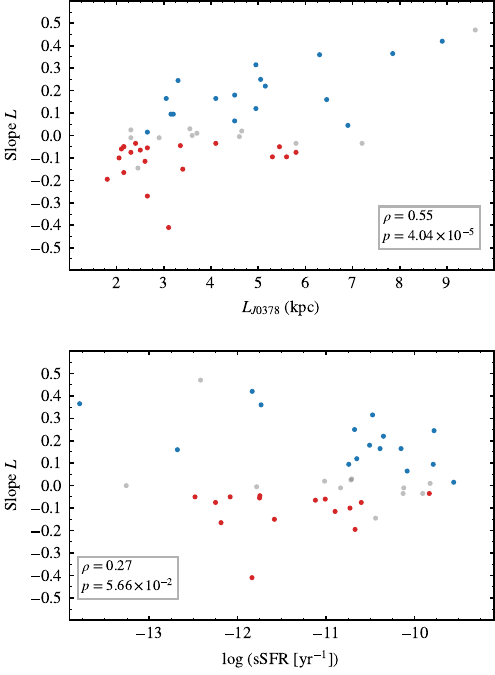}
    \caption{Two‐panel comparison of the fitted slope of bar‐length measurements across bands for our 50‐galaxy sample. Top panel: slope versus bar length measured in the \textit{J0378} filter ($L_{J0378}$). Bottom panel: slope versus specific star formation rate ($\log\,\mathrm{sSFR}/\mathrm{yr}^{-1}$). In both panels, blue points denote galaxies with a strong positive monotonic trend, red points denote those with a strong negative trend, and gray points show the remainder. The Spearman correlation shown refers to the full sample.}  
        \label{fig:slope_L_L}
\end{figure}

Fig.~\ref{fig:slope_L_L} presents a two‐panel comparison of the fitted slope of bar‐length variation across bands against two key parameters, revealing bimodality. In the upper panel, the slope is plotted versus the bar length measured in the $J0378$ filter ($L_{J0378}$): galaxies with shorter $L_{J0378}$ (< 4 kpc) predominantly exhibit negative monotonicity, meaning their bars appear systematically larger in redder filters, whereas those with longer $L_{J0378}$ show positive monotonicity, with bars systematically larger in bluer filters. This behaviour arises because the $J0378$ filter preferentially traces star‐forming regions in the bar, so intrinsically longer bars in this filter, hosting more star formation, display a stronger wavelength‐dependent lengthening. In the lower panel, the same slope is compared to sSFR, again separating the sample into two subgroups: galaxies with $\log(\mathrm{sSFR}/\mathrm{yr}^{-1})<-11$ preferentially exhibit negative monotonicity ($n=20$), while those with $\log(\mathrm{sSFR}/\mathrm{yr}^{-1})\ge-11$ more often positive monotonicity ($n=18$). This split highlights two distinct galaxy populations, defined by their star‐formation activity, that exhibit opposite wavelength‐dependent bar‐length trends. The Spearman correlation coefficient shown in each panel refers to the full 50-galaxy sample.

\begin{figure*}
	\includegraphics[]{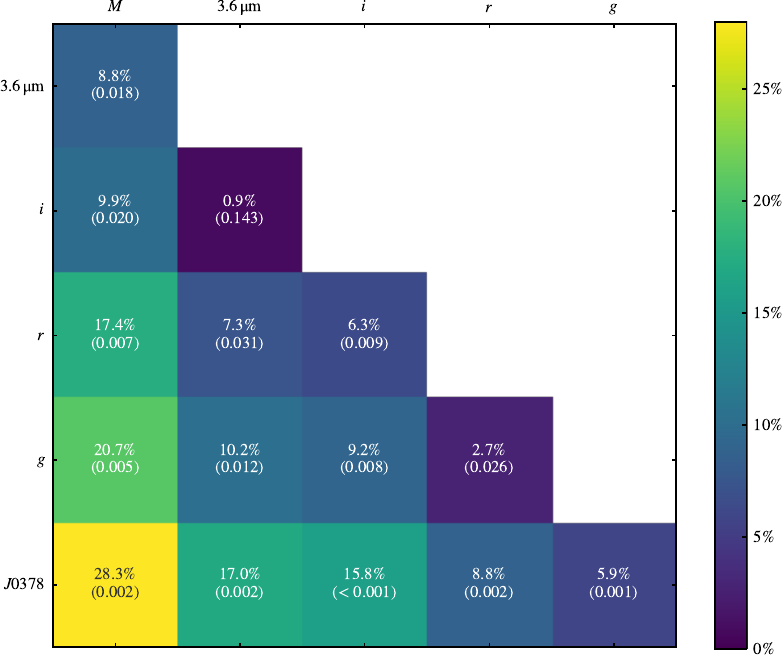}
    \caption{Comparison of bar lengths across different bands for the star-forming subsample ($\log(\text{sSFR}) \geq -11~\mathrm{yr}^{-1}$). The percentages indicate how much longer the bars appear on average when comparing the filters. The statistical significance of each measurement was assessed using the Wilcoxon test, with the corresponding $p$-values shown in parentheses.}
    \label{fig:meanratio_L_sSFR11}
\end{figure*}

Based on the interval of $\log(\text{sSFR}) \ge -11$, we can separate the two samples presented in Fig.\ref{fig:hist_rho_SMA} and reevaluate the bar lengths across different filters. The outcome of this star-forming subsample analysis is illustrated in Fig.\ref{fig:meanratio_L_sSFR11}. Unlike the previously inconclusive results, a clear trend now emerges: selecting only galaxies with a specific star formation rate of $\log(\text{sSFR}) \ge -11~(\mathrm{yr}^{-1})$ reveals that barred galaxies are systematically longer in bluer filters, as shown in Fig.\ref{fig:meanratio_L_sSFR11}. This distinction between the two galaxy populations can be further observed by comparing Fig.\ref{fig:hist_rho_SMA} with Fig.~\ref{fig:slope_L_L}.

\subsection{Relation between stellar ages and waveband dependence}

In the previous subsections, we quantitatively characterized the effects of different wavebands on the properties of simulated bars. In this final subsection, we use the ages of stellar populations in the simulation to investigate a possible physical mechanism behind the detected dependencies.

We selected stellar particles into three distinct groups based on their ages: young ($\leq$ 2.0 Gyr), intermediate (2 < age < 8 Gyr), and old ($\geq$ 8 Gyr). Using this selection, we generated mass maps following the same methodology as before. Then, we applied the ellipse-fitting method to determine the length and ellipticity of the bars visible in each map.

\begin{figure}
	\includegraphics[]{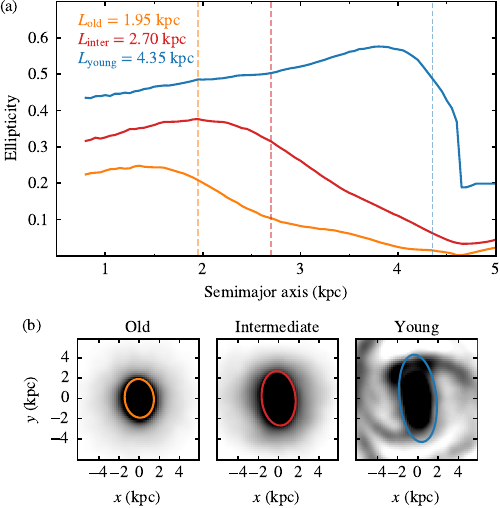}
    \caption{(a) Results of the ellipse-fit on the mass maps for different stellar age groups. The solid lines represent the fits for old ($\geq 8$ Gyr, orange), intermediate ($2 < \text{age} < 8$ Gyr, red), and young ($\leq 2.0$ Gyr, blue) stellar populations. The dashed lines indicate the estimated bar length in each case. (b) Spatial representation of the ellipses obtained from the fits, illustrating their bar size and ellipticity for each stellar age group. The panels show the fitted ellipses over the mass maps for old (left), intermediate (middle), and young (right) populations. The barred galaxy used in this example is the subhalo ID548418.
    }
    \label{fig:ages_massmaps}
\end{figure}

Similarly to the results obtained with the creation of mock images, which simulate different observational bands sensitive to distinct stellar populations, the mass maps filtered by age also reveal a trend: older stars form shorter and less elliptical bars, while younger stars result in longer and more elliptical bars (Fig. \ref{fig:ages_massmaps}). This pattern is particularly evident in the rightmost panel of the lower frame (b) in Fig. \ref{fig:ages_massmaps}, where the bar appears significantly elongated, with spiral arms emerging from its ends and a distinct clump of particles aggregating at the bar extremity—features absent in the mass map of older stars. Additionally, the old population frame exhibits a much rounder structure.

The lack of young stellar particles plays a key role in this analysis. In galaxies that have undergone quenching or exhibit low star formation rates, the reduced presence of young stars in the bar region hinders the selection process, making it impossible to generate a sufficiently dense mass map. In some cases, particularly in systems with very low star formation, the bar itself may become unidentifiable, resulting in a mass map that resembles a UV mock image, which exhibits a similar behaviour.

\section{Discussion}
    \label{sec:discussion}

\subsection{Bar properties are waveband-dependent in TNG50}

The analysis of the mock images across different observational filters suggests that star formation regions in the galaxies of our sample may exhibit noticeable variations depending on the observed wavelength. These phenomena dependent on stellar populations can be simulated through realistic mock imaging techniques; \citet{2024Baes_AtlasTNGSKIRT_radius} has already explored a wavelength-dependent effect within IllustrisTNG, though in a different context, where the authors explores the wavelength dependence of the effective radius across multiple spectral bands (\textit{u, g, r, i, z, J, H, Ks}). 

The ellipse fitting method allows us to quantify these differences in a region where different stellar populations are present, in the galactic bar. Comparing these measurements highlights two primary characteristics of the bars: their length and ellipticity. The comparison across different filters highlights a strong wavelength dependency in bar ellipticity, with a clear trend where bluer filters (Fig.~\ref{fig:corner_plot_e_L}) exhibit higher ellipticity than their longer-wavelength counterparts. The magnitude of this difference indicates that the largest ellipticity discrepancy between filters can reach 15 per cent, comparing the blue SPLUS \textit{J0378} filter to the stellar mass maps, \textit{M} (Fig.~\ref{fig:meanratio_E}). The smallest difference, at 1.3 per cent, is found between the \textit{r} and \textit{i} filter.

The bar length is also influenced by the observed waveband; however, its behaviour differs from that of ellipticity. If we observe the entire sample (Fig. \ref{fig:meanratio_L}), the average difference in measurement between the filters is small and lacks statistical significance according to the Wilcoxon test. This behaviour is not caused by the bar having the same length across filters, but rather by two distinct galaxy samples, separated by their star formation activity, exhibiting opposing behaviours: one with the bar appearing systematically longer in bluer filters, and the other appearing systematically longer in redder filters. The average of these values results in differences close to zero. This population can be understood by observing Fig.\ref{fig:hist_rho_SMA} and Fig.\ref{fig:slope_L_L}. When separating the sample by selecting galaxies with $\log(\text{sSFR}/\mathrm{yr}^{-1}) \geq -11~$, there is a clear trend with a significant difference between the filters and statistical significance (Fig.\ref{fig:meanratio_L_sSFR11}): bars are statistically longer in bluer filters.

The opposite behaviour observed in galaxies with low star formation rates, where bars appear longer in redder filters, can be explained by the absence of star-forming regions (e.g., clumps near the ends of the bar). In galaxies with high sSFR, these regions contribute significantly to the apparent bar extent in blue filters such as $g$ and $J0378$, resulting in longer measured bars. In contrast, passive galaxies do not exhibit such regions, which reduces bar prominence in blue bands and leads to more consistent bar lengths across filters. This behaviour is consistent with findings from stellar population studies showing that bars host, on average, older and more metal-rich stars than the surrounding disk \citep{2020Neumann_bar_pop, 2021Bittner} and that younger stars, when present, tend to be found near the ends of the bar due to kinematic fractionation \citep{2017Athanassoula_kinematic_MW_bulge, 2017Fragkoud_kinematic_frac_bulge, 2017Debattista_bar_kinematic_pop}. In the absence of such young populations, infrared observations (e.g., at $3.6 , \upmu$m) trace the older stellar population and the underlying mass distribution more reliably, offering a clearer picture of the full bar structure compared to optical bands, which are more affected by dust extinction and star formation biases \citep{2000Eskridge_barfrac_infrared, 2018Erwin_bardfrac_local}. These effects are incorporated in our RT-based mocks through dust and stellar population modeling. Furthermore, in galaxies with negative Spearman correlations between bar length and wavelength, the average absolute slope is approximately half that measured in galaxies with $\log(\mathrm{sSFR}) \ge -11$. This suggests that the bar-length difference across filters in passive systems is smaller when compared to the analogous effect seen in actively star-forming galaxies.

\subsection{Mass maps are similar to infrared filters}

In the context of galaxy simulations and observations, comparisons are commonly based on mass distribution. However, an accurate translation from light to mass is crucial, as observational studies often rely on this approach to infer stellar and dust properties.

Taking this aspect into consideration, one of the most important results of our paper is the comparison between measurements of bar properties in stellar mass maps and in observational filters, the latter resulting from radiative transfer mocks. We emphasize that in the context of cosmological simulations, comparing the filters in relation to the mass distribution has particularly useful results, in some cases more useful than comparing the filters with each other. This result is illustrated by the solid black line in Fig.~\ref{fig:ellipse_fitting_filters}, which shows that the mass map yields bar ellipticity and length values similar to those derived from infrared filters, particularly the \textit{Spitzer} IRAC/$3.6 , \upmu \text{m}$ channel. However, the ellipticity profiles derived from the mock filters exhibit distinct features that are not present in the mass map. This emphasizes that the observed isophotal shapes in infrared bands are not determined solely by the underlying mass distribution, but also reflect the effects of stellar population variations and the interaction of light with dust, especially in high-density regions such as galaxy centres and the connections between the bar and the spiral arms. These features arise from radiative transfer effects in the mock images, which incorporate both the spatial distribution of stellar light and its attenuation by dust.

This similarity is to be expected, as infrared wavelengths are less affected by extinction and more accurately trace the large population of low-mass stars that dominate the galactic disc \citep{2000Eskridge_barfrac_infrared}. These stars, although individually less massive, collectively dominate the galactic disc due to the distributions prescribed by the Initial Mass Function (IMF), as detailed in \citet{2003Chabrier_massfuction}.  Unlike blue filters, which prominently display young stellar regions with active star formation, the infrared view focuses on the broader, underlying mass distribution, where such actively star-forming regions are less emphasized.

For these reasons, the infrared bands are generally considered the closest proxy to the mass maps. Nevertheless, our measurements indicate that simulated bars measured using mass maps will appear 10 per cent shorter, compared to the same measurement done in the $3.6 \, \upmu \text{m}$ mock.

These results are relevant, as they can guide both future studies and interpretations of previously published results by the galaxy hydrodynamic simulation community. When compared with observational results, the simple mass distribution will generally resemble infrared filters; however, it may diverge considerably from results in bluer filters, in the optical and UV.

\subsection{Bars are not often detectable in UV mocks}
    \label{sec:UV_bars}

In Appendix \ref{fig:appendix1_A}, it can be noted that the bars may disappear or be attenuated in ultraviolet filters. This aspect prevented measurements dependent on ellipse fitting convergence from being performed for the \textit{Galex FUV} and \textit{NUV} filters in our sample. Observational studies also report difficulties in measuring the bar in UV filters \citep{2003Sheth_weakbars_identific, 2024Karin_wavebandependence}.

The bar disappears in over 90 per cent of our sample, primarily for two reasons. Firstly, this is expected as bars in massive galaxies tend to suppress star formation within the bar region itself, leading to a lack of very young, massive stars \citep[e.g.][]{2020Diaz-Garcia_GALEX_bars, 2020MNRASFraser_MANGA_bars_starformation, 2023Maeda_star_formation}. Consequently, there is an insignificant amount of emission by the dominant bar stellar population at these ultraviolet wavelengths, except potentially at the bar ends and center. Secondly, the bar might not be distinguishable from other star-forming clumps observed at this wavelength. In the latter case, the ellipse fit method does not prove to be an effective method to assess the presence of a bar in the ultraviolet.

\subsection{Realism in mock images is crucial when fitting ellipses}
We highlight that the realistic noise addition step is essential both for reproducing these results and for applying similar techniques. In the process of measuring bars through ellipse fitting, we encountered particularities in the isophotes of simulated galaxies \citep{2024Goncalves_PSF}, whose results differed from the expected shapes observed in real galaxies. In the simulations, the central isophotes of the bar tend to become excessively elongated towards the center of the galaxy, and this behavior tends to hinder the identification of the ellipticity peak that marks the end of the bar. However, the analogous behavior of the isophotes of real galaxies is reproduced after simple noise addition steps, such as PSF convolution and the simulation of synthetic noise, including sky background and read noise, as described in the methodology section. This addition of noise ends up attenuating the central ellipticity, which makes the ellipticity profile compatible with the observations.

A similar issue regarding the excessive elongation of central isophotes was noted by \citet{2024Lu_ellipse_TNG50}, who found that ellipticity profiles in simulated TNG50 galaxies tend to increase towards the centre. To mitigate this problem, they proposed starting the measurement of bar properties from a boundary limit of 1.4 kpc, as shown in their figure~1. This approach prevents the severe underestimation of bar lengths that occurs when using methods such as those related to the maximum ellipticity or 85 per cent of the maximum ellipticity, which, when applied from the centre, lead to excessively short bars. However, this choice introduces a bias, as the location of the ellipticity peak, and therefore the measured bar length, becomes dependent on the arbitrary choice of the inner limit. In contrast, our solution, which involves applying a PSF sufficient to correct the central isophotes, resolves this issue naturally. By accounting for the instrumental effects of real observations, our method restores the expected ellipticity growth towards a maximum, rather than towards the galaxy centre, ensuring a more robust and unbiased measurement of bar properties.

The influence of PSF addition on bar length measurement has also been highlighted in previous studies. \citet{2022Frankel_TNG_barpattern} demonstrated that applying a typical SDSS-\textit{r} PSF to simulated images from TNG50 can increase the estimated bar length when measured using Fourier coefficients. Figure~2 of their work shows that the bar extent derived from the stellar mass distribution is consistently shorter than the measurements obtained from PSF-convolved images. Although we use an ellipse-fitting method to determine the bar, this effect underscores the impact of the PSF on the ellipticity of the bar. The smoothing of the bar structure induced by the PSF can impact the location of the ellipticity peak, directly affecting the determination of the bar length.

In the context of our analysis, the addition of synthetic noise, including PSF convolution and background noises, has been shown to significantly impact the measurement of structural properties, such as bar length and ellipticity profiles.

\subsection{Waveband-dependence is intrinsic to the host galaxy}

Our statistical analysis reveals a clear waveband dependency for a subset of our galaxies: the bluer the filter, the thinner and longer barred galaxies appear in our observations. These measurements are consistently thinner and longer than those derived from the stellar mass map. However, it is important to note that some galaxies do not exhibit this dependence, displaying  the same bar measurement regardless of the filter used. Thus, the effect cannot be solely attributed to differences in measurements across filters, as some galaxies follow the trend while others do not. Instead, the trend that generally appears more pronounced in galaxies with active star formation implies that intrinsic properties such as stellar populations and their activity significantly influence the measured bar properties. This suggests that the phenomenon is tied to specific properties of the stellar populations or to the presence of a morphological component that enables or inhibits this wavelength dependence of the bar. 

We find that galaxies in our sample show greater differences between filters when they have a high specific star formation rate. This analysis was made possible by measuring the slope of the difference between filters. We highlight that this trend is even stronger when considering only the strictly waveband-dependent galaxies.

This evidence suggests the existence of a morphological component that activates or inhibits the wavelength dependence of bar properties. Currently, \citet{2024Karin_wavebandependence} proposes two hypotheses to explain this phenomenon, also observed in real galaxies: the presence of knots of star formation at the ends of the bar or the effect of kinematic fractionation \citep{2017Athanassoula_kinematic_MW_bulge, 2017Fragkoud_kinematic_frac_bulge, 2017Debattista_bar_kinematic_pop, 2020Neumann_bar_pop}, which would make the bar appear more elongated and narrow in blue filters; or the presence of a rounder structure overlapping, composed of old stars, whose overlapping isophotes would make bar measurements less elliptical in redder filters, with a reduced effect in blue filters, thereby elongating and making the bar more elliptical.

Now that we have confirmed this wavelength-dependence phenomenon within IllustrisTNG, the tools of this simulation will be valuable for attempting to isolate the component responsible for this behaviour.

\section{Summary and Conclusion}
    \label{sec:conclusion}

We created realistic mock images of 50 barred galaxies from TNG50, using filters from infrared to UV, and measured the bar properties of this sample using ellipse fitting. The main results of this paper can be summarized as follows:

\begin{enumerate}
    \item Bars appears systematically thinner in bluer filters for over three-quarters of the sample. This trend is noted when measuring ellipticity across all filters ($3.6 \, \upmu \text{m}$, $i$, $r$, $g$, and J0378), with ellipticity increasing by 6 percent and 9 percent respectively from the $3.6 \, \upmu \text{m}$ to the $g$ and $J0378$ bands. These findings are consistent with \citet{2024Karin_wavebandependence}.

    \item For bar length, the mean difference across filters is inconclusive when regarding the entire sample. This is due to passive galaxies showing shorter bars in bluer filters (e.g. $J0378$ and $g$),  which drives the inverted trend, with an average slope approximately half that of the star-forming group, meaning the bar-length difference across filters is smaller. When only star-forming galaxies are considered, then we find a statistically significant mean difference in bar length across filters. Bar length, in this star-forming subsample, increases by 10 per cent and 17 per cent from the $3.6 \, \upmu \text{m}$ to the $g$ and $J0378$ bands, respectively, and this behaviour aligns with the findings of \citet{2024Karin_wavebandependence}.

    \item Bars disappear in over 90 per cent of our UV sample, primarily because the bars dominant stellar populations emit minimally at UV wavelengths and are often indistinguishable from surrounding star-forming regions. In the few cases where bars remain, this complicates bar detection using ellipse fitting methods. These challenges are consistent with findings from studies such as those by \citet{2003Sheth_weakbars_identific} and \citet{2024Karin_wavebandependence}, which highlight difficulties in detecting bars in UV observations.

    \item Wavelength dependence is stronger in galaxies with higher specific star formation rates, indicating that young star-forming regions may significantly drive the observed differences between filters. This effect is evident for the entire sample when analysing ellipticity, but for  bar length, it is revealed only when regarding separately the galaxies in the star-forming subsample.

    \item This wavelength dependence supports a morphological hypothesis: star-forming clumps at the bar ends, linked to spiral arms, could extend and elongate it in bluer filters.
    
    \item Mass maps exhibit smoother profiles and tend to show bars as shorter and less elliptical compared to filters, reflecting the dominant influence of the older, low-mass stellar population that constitutes the bulk of stellar mass in these structures. Nevertheless, our measurements indicate that simulated bars measured using mass maps will appear 10 per cent shorter, compared to the same measurement done in the $3.6 \, \upmu \text{m}$ mock. For the star-forming subsample, the bar can appear up to 20 -- 30 percent longer in bluer mocks than in the mass map.

    \item Mass maps filtered by age reveal a clear trend: older stars form shorter, rounder bars, while younger stars generate longer, more elliptical ones, confirming previous findings in the literature \citep{2020Neumann_bar_pop}. This effect depends on the galaxy having sufficient number of young stellar particles, making it less noticeable in galaxies with low star formation or undergoing quenching. The presence of spiral arms and clumps at the bar extremity in younger populations, contrasts with the older ones, whose distribution resembles a rounder less elliptical isophote.

\end{enumerate}

In conclusion, we suggest expanding this analysis, currently done at \(z=0\), to higher redshifts, which could help identify the galaxy component that inhibits or enables the waveband dependence of bar properties. Furthermore, the use of realistic mock images proved to be a promising approach, functioning as a bridge between observational results and simulated galaxies.

\section*{Acknowledgements}
We thank the anonymous referee for the helpful suggestions which contributed to improving the paper.
GFG acknowledges support from the Brazilian agency Conselho Nacional de Desenvolvimento Científico e Tecnológico (CNPq), Fundação Araucária, and the IllustrisTNG collaboration for generously providing access to TNG50 simulation data and computational resources via the online JupyterLab workspace. REGM acknowledges support from CNPq, through grants 406908/2018-4 and 307205/2021-5, and from Fundação Araucária through grant PDI 346/2024 -- NAPI Fenômenos Extremos do Universo. KMD thanks the support of the Serrapilheira Institute (grant Serra-1709-17357) as well as that of the Brazilian National Research Council (CNPq grant 308584/2022-8) and of the Rio de Janeiro Research Foundation (FAPERJ grant E-32/200.952/2022), Brazil.

\section*{Data Availability}

The raw data supporting the conclusions of this article will be made available upon reasonable request to the corresponding author.



\bibliographystyle{mnras}
\bibliography{mock_bars} 



\appendix

\section{PSF Sensitivity Analysis}

\begin{figure}
    \centering
    \includegraphics[width=\linewidth]{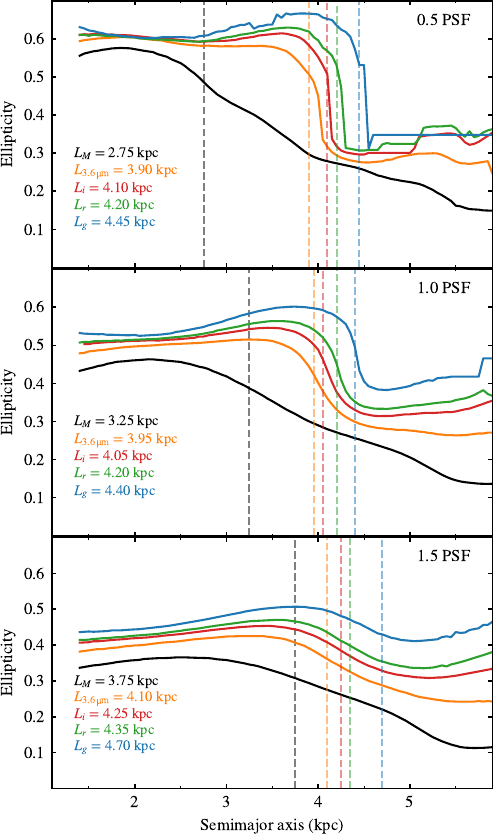}
    \caption{The panels demonstrate the impact of varying the PSF on the profiles, presenting results for 0.5 PSF (top), 1.0 PSF (middle), and 1.5 PSF (bottom). These are ellipticity profiles as a function of the semimajor axis, resulting from ellipse fits to the isophotes. The solid lines represent the profiles for the  Mass (black), \textit{Spitzer} $3.6 \, \upmu \text{m}$ (orange), SDSS \textit{r} (red), SDSS \textit{i} (green) and SDSS \textit{g} (blue). The estimated bar lengths for these bands are indicated in each panel.}
    \label{fig:appendix_B}
\end{figure}

This section details the influence of the Point Spread Function (PSF) choice on our morphological analysis, with reference to Fig.~\ref{fig:appendix_B}. In the main analysis of this study, we employed a standard PSF with a FWHM of ~$\sim$~1 kpc. This value corresponds to approximately ~$\sim$~1 arcsecond for galaxies at a redshift of $z$~$\sim$~0.05. The smoothing of the images using this PSF was performed using a Gaussian convolution algorithm, available in the Astropy library \citep{2018_astropy1}. The results obtained with this reference PSF are presented in the middle panel.

Fig.~\ref{fig:appendix_B} further explores the impact of PSF variation on the results. For this purpose, analyses are presented using a smaller PSF (FWHM of 0.5 kpc, labeled ``0.5 PSF'' in the top panel) and a larger PSF (FWHM of 1.5 kpc, labeled ``1.5 PSF'' in the bottom panel). This comparison is useful to assess the robustness of the conclusions. It is recognized that the choice of PSF can influence quantitative measurements, such as maximum ellipticity values  and, consequently, estimates of bar length. The smoothing process inherent in convolution with the PSF can alter the apparent strength of the bar, and in scenarios with greater smoothing, intrinsically weaker or shorter bars may become difficult to detect or have their characteristics underestimated.

Despite the sensitivity of absolute metrics to the PSF, Fig.~\ref{fig:appendix_B} demonstrates that crucial aspects of the ellipticity profiles are preserved. Notably, the figure shows that the relative ordering of maximum ellipticities and estimated bar lengths for the different spectral bands remains consistent across the investigated PSF variations. This consistency is particularly important, suggesting that the monotonicity phenomenon reported in our study is a robust characteristic of the galaxies data and is not dependent on the specific PSF choice.

The SPLUS J0378 filter was not included in the comparative examples. This omission is due to technical difficulties encountered during the ellipse fitting process for this particular filter close to the UV.

\section{Mock Images of the Galaxy Sample}
We show in Fig. \ref{fig:appendix1_A} mock images of the galaxy sample used in our analysis. These images were generated using synthetic radiative transfer simulations performed with SKIRT and are presented in their noise-free version. The mock observations span multiple wavebands filters, including \textit{Spitzer} $3.6 \, \upmu \text{m}$, SDSS \textit{i}, SDSS \textit{r}, SDSS \textit{g}, SPLUS \textit{J0378}, \textit{GALEX NUV}, and \textit{GALEX FUV}. Each column corresponds to a different filter, presenting various morphological and structural features of the galaxies.

These mock images serve to illustrate how the galaxies appear at different wavelengths, providing insights into their stellar populations, dust content, and star formation regions. Of particular interest is the appearance of the bar structure and its properties, which are clearly identifiable in the infrared and optical bands but becomes indistinguishable in the UV due to the dominance of young star-forming regions. The images are organized in sequential figures, maintaining the same subhalo ID as in the TNG50 catalogs.

\begin{figure*}
    \centering
    \includegraphics[width=\linewidth]{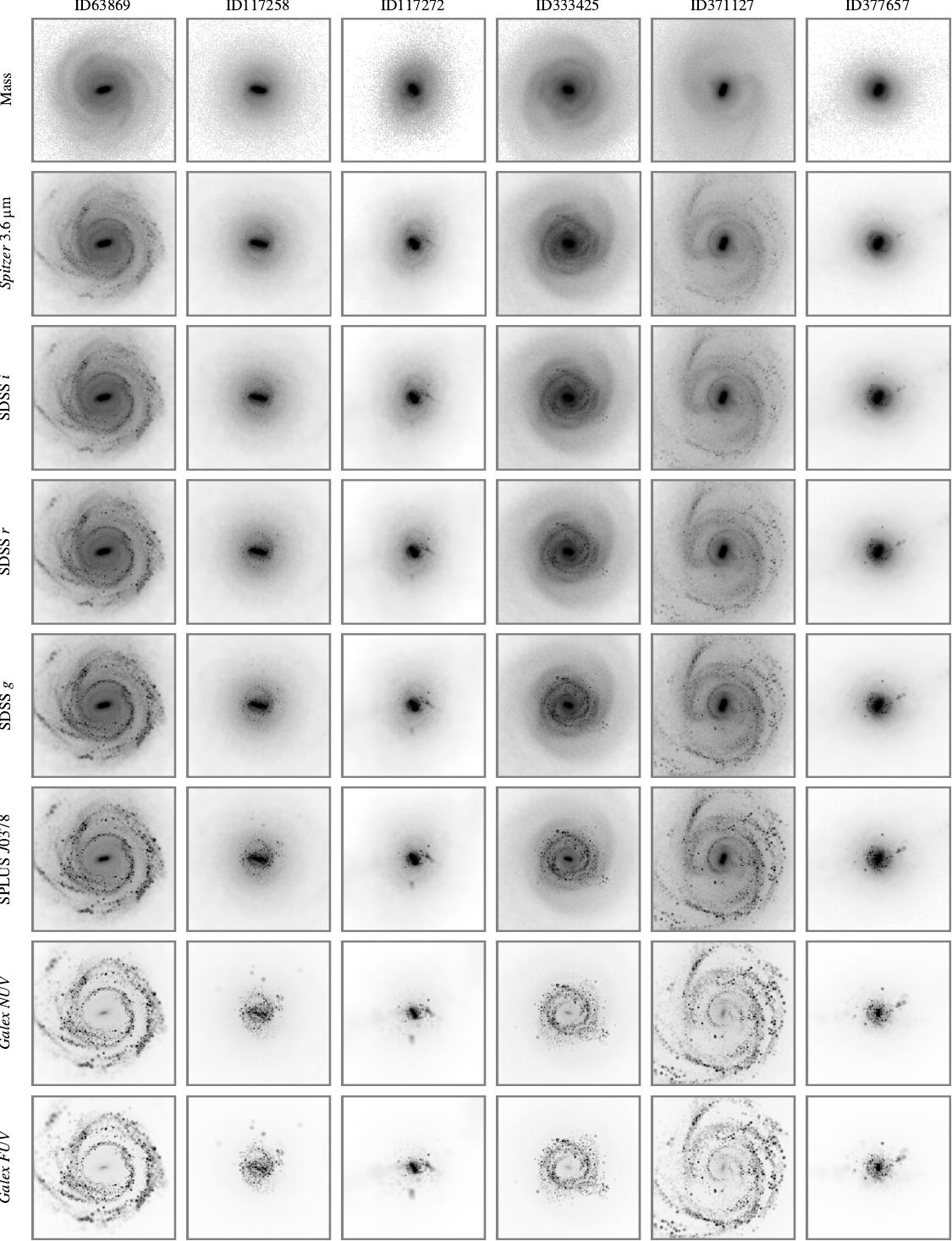}
    \caption{Mass maps and mock images of our galaxy sample in different wavebands: \textit{Spitzer} IRAC/$3.6 \, \upmu \text{m}$, SDSS \textit{i}, SDSS \textit{r}, SDSS \textit{g}, SPLUS \textit{J0378}, \textit{Galex NUV}, and \textit{Galex FUV}. All the figures were generated with a size of 60 $\times$ 60 kpc in each frame and with a face-on orientation.}
    \label{fig:appendix1_A}
\end{figure*}

\begin{figure*}\ContinuedFloat
    \centering
    \includegraphics[width=\linewidth]{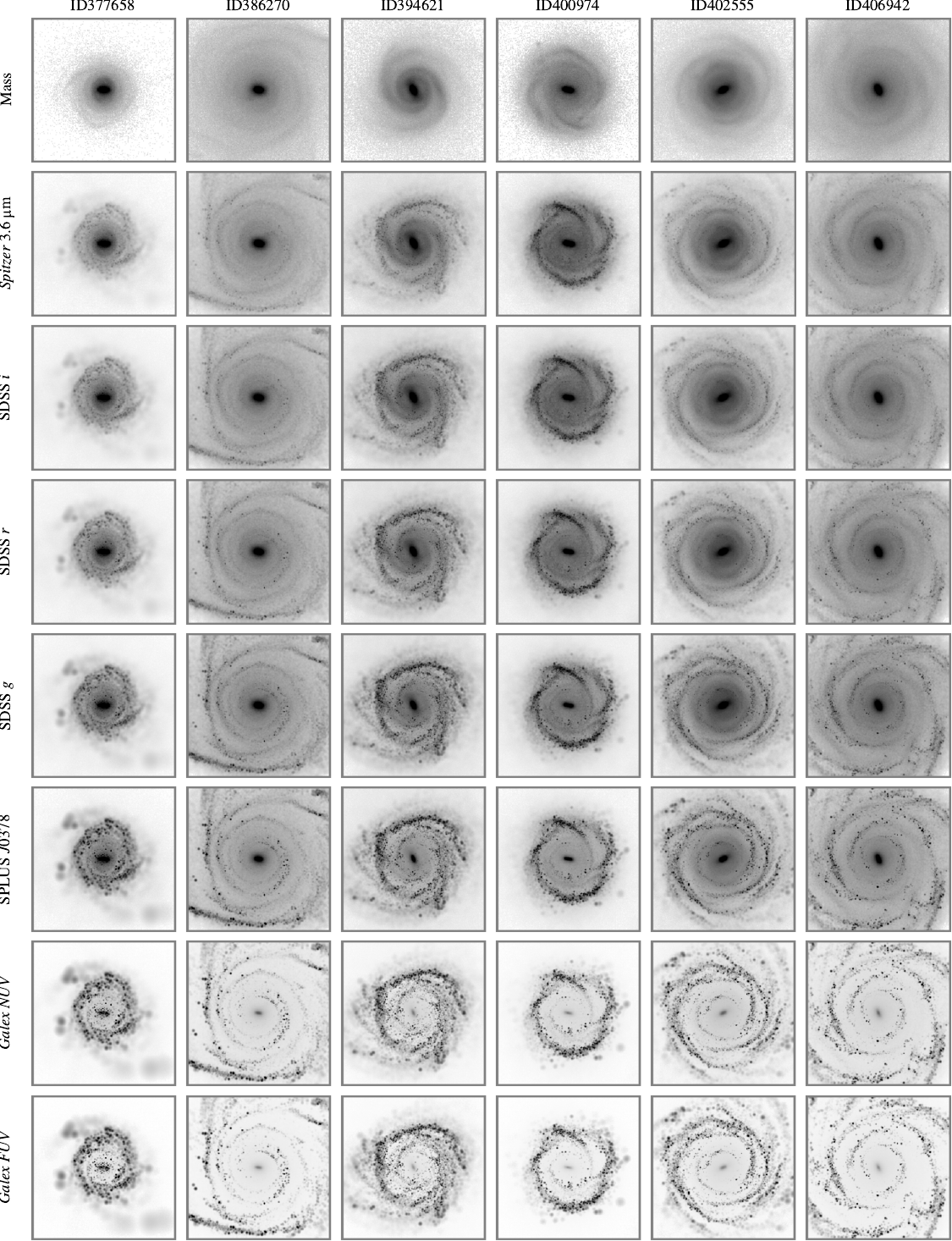}
    \caption{Continued.}
    \label{fig:appendix1_B}
\end{figure*}

\begin{figure*}\ContinuedFloat
    \centering
    \includegraphics[width=\linewidth]{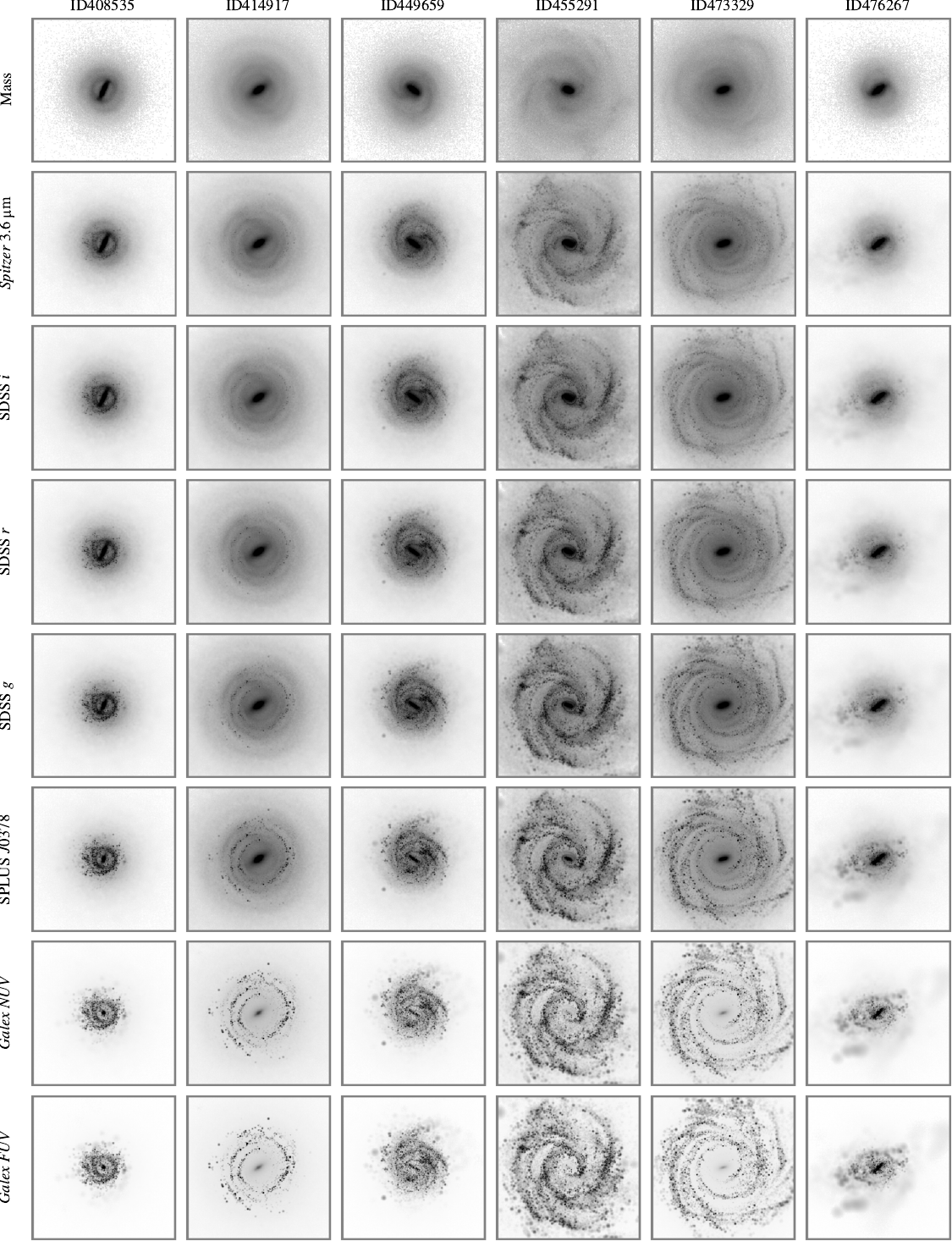}
    \caption{Continued.}
    \label{fig:appendix1_C}
\end{figure*}

\begin{figure*}\ContinuedFloat
    \centering
    \includegraphics[width=\linewidth]{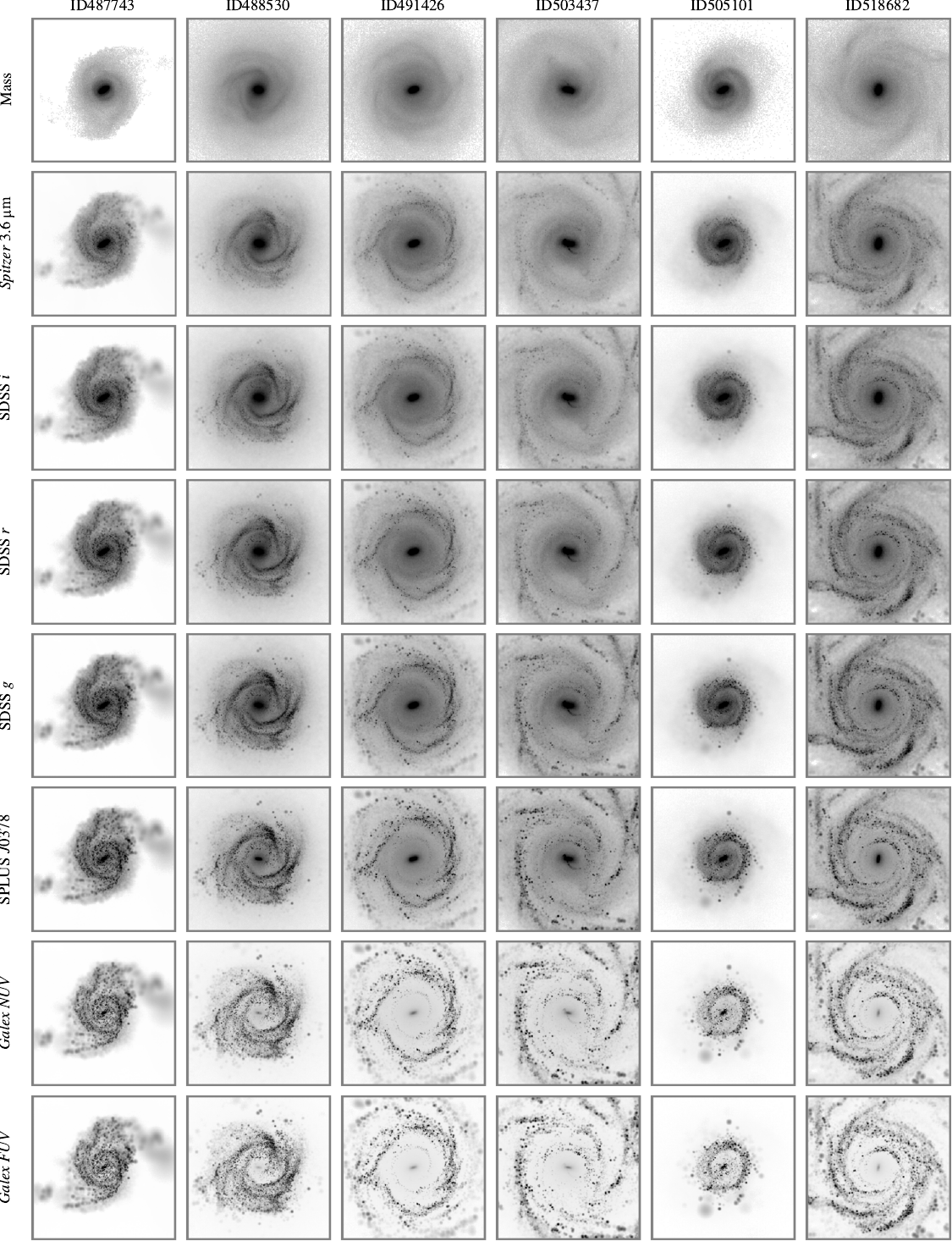}
    \caption{Continued.}
    \label{fig:appendix1_D}
\end{figure*}

\begin{figure*}\ContinuedFloat
    \centering
    \includegraphics[width=\linewidth]{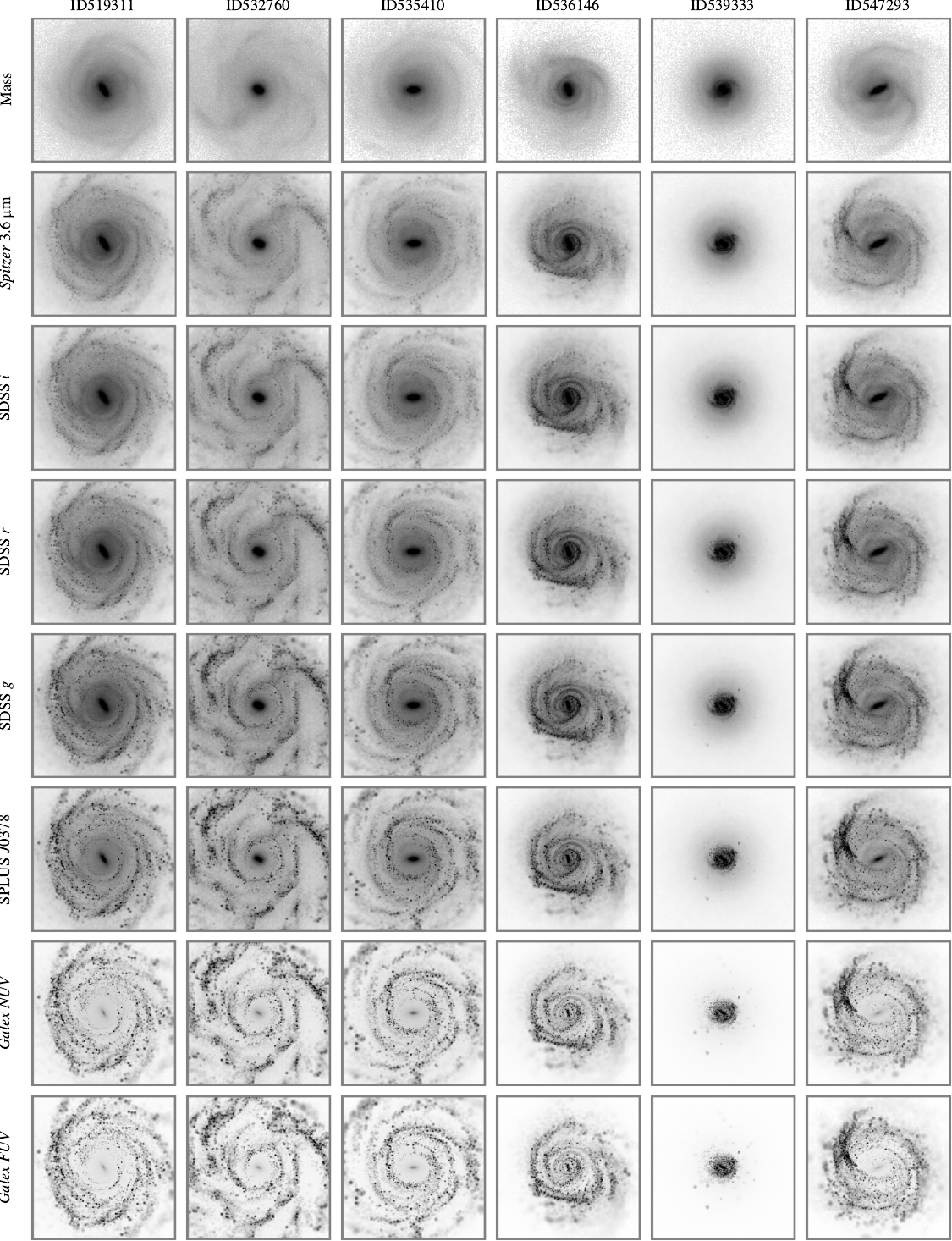}
    \caption{Continued.}
    \label{fig:appendix1_E}
\end{figure*}

\begin{figure*}\ContinuedFloat
    \centering
    \includegraphics[width=\linewidth]{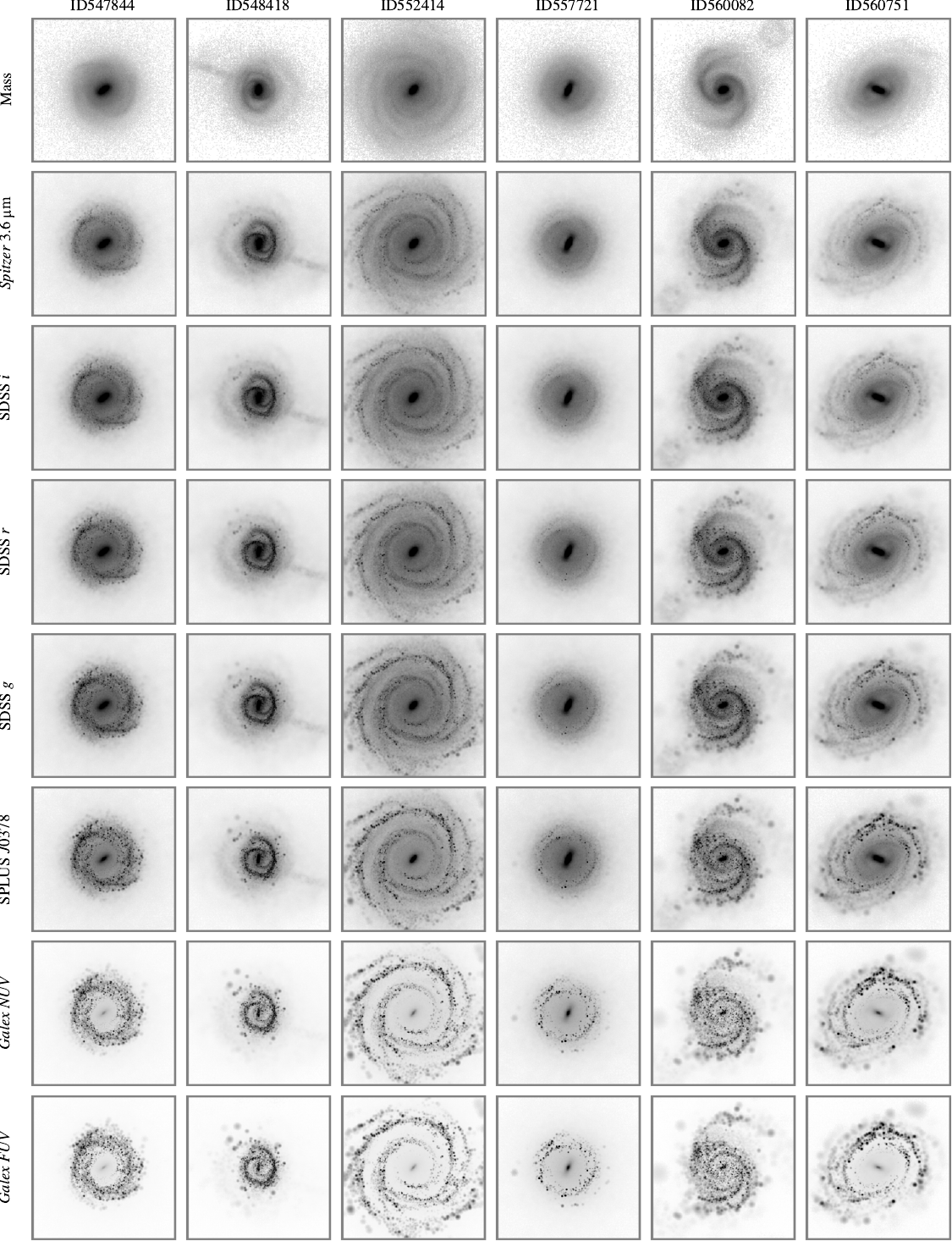}
    \caption{Continued.}
    \label{fig:appendix1_F}
\end{figure*}

\begin{figure*}\ContinuedFloat
    \centering
    \includegraphics[width=\linewidth]{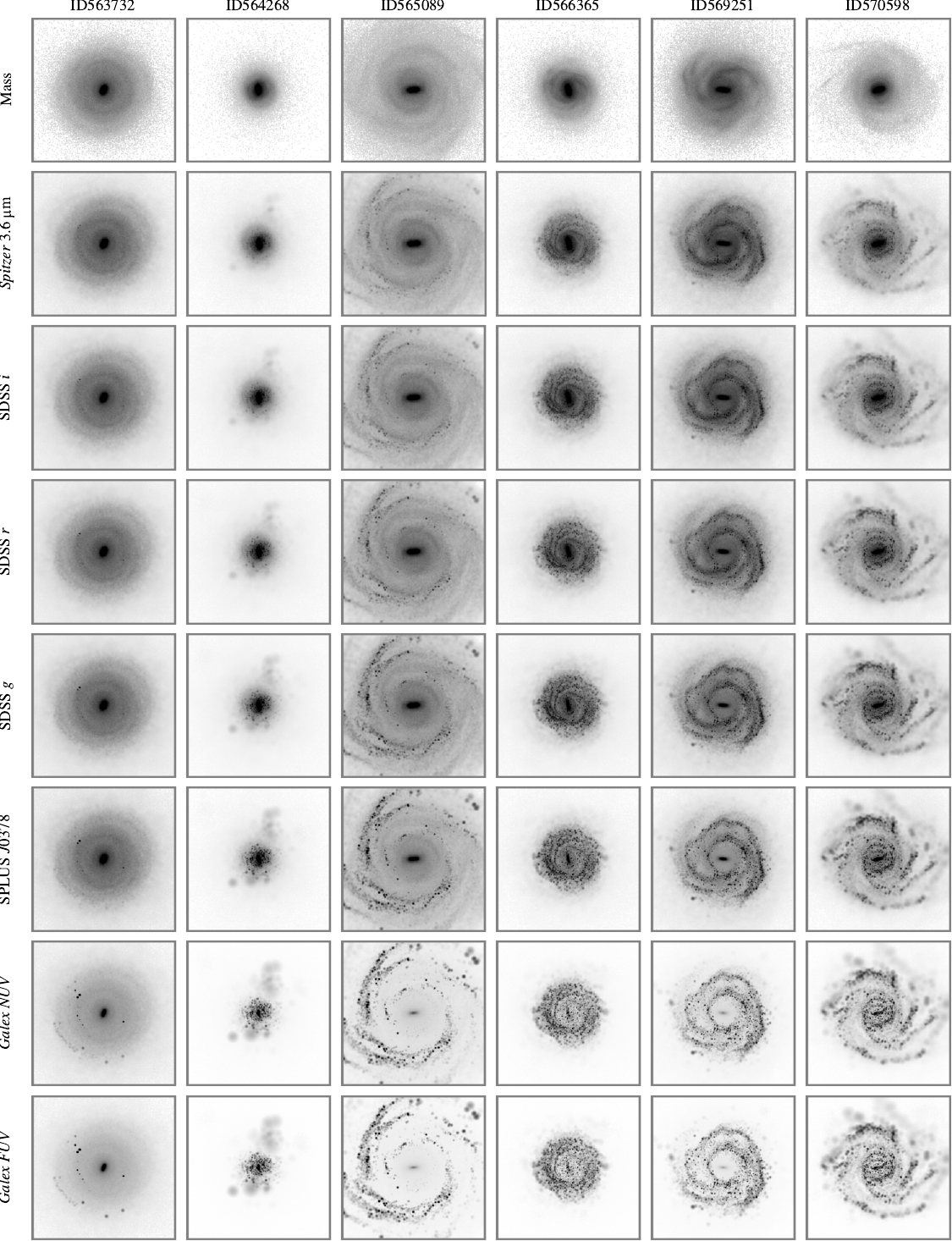}
    \caption{Continued.}
    \label{fig:appendix1_G}
\end{figure*}

\begin{figure*}\ContinuedFloat
    \centering
    \includegraphics[width=\linewidth]{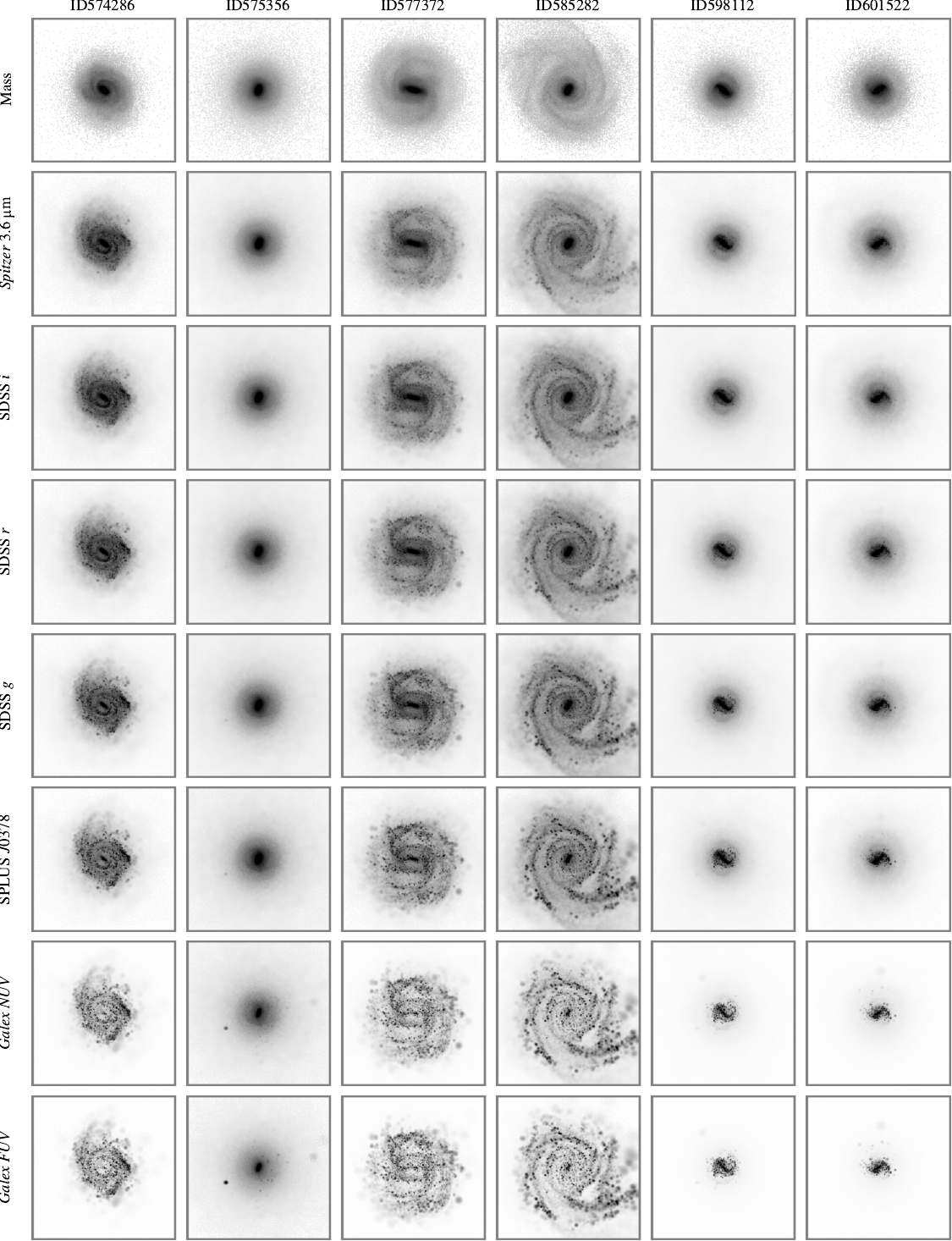}
    \caption{Continued.}
    \label{fig:appendix1_H}
\end{figure*}

\begin{figure*}\ContinuedFloat
    \centering
    \includegraphics[height=1.0\textheight]{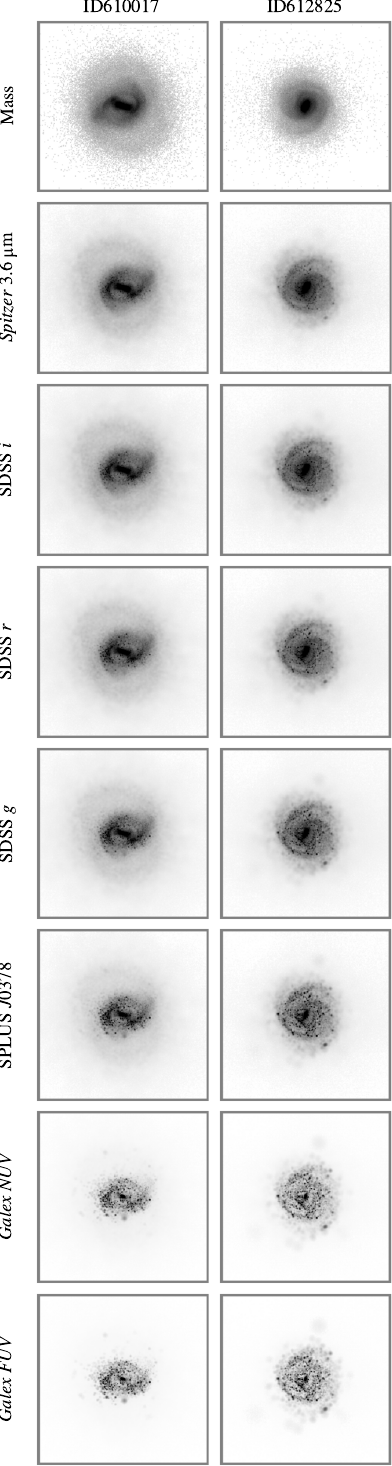}
    \caption{Continued.}
    \label{fig:appendix1_I}
\end{figure*}

\bsp	
\label{lastpage}
\end{document}